\documentclass[english,pra,floatfix,superscriptaddress,twocolumn,showpacs,amsmath,amssymb,reprint,notitlepage]{revtex4-1}
\usepackage[T1]{fontenc}
\usepackage[latin9]{inputenc}
\usepackage{color}
\usepackage{amsmath}
\usepackage{amssymb}

\makeatletter
\PassOptionsToPackage{caption=false}{subfig} 
\usepackage{hyperref}
\hypersetup{
breaklinks=true,
colorlinks=true,
citecolor=blue,
linkcolor=blue,
filecolor=blue,
urlcolor=blue
}
\numberwithin{equation}{section}

\IfFileExists{lmodern.sty}{\usepackage{lmodern}}{}
\makeatother

\usepackage{babel}
\begin{document}

\title{Electromagnetic response of superconductors in the presence of multiple
collective modes}

\author{Rufus Boyack}

\affiliation{Department of Physics \& Theoretical Physics Institute, University
of Alberta, Edmonton, Alberta T6G 2E1, Canada}

\author{Pedro L. S. Lopes}

\affiliation{Department of Physics and Astronomy \& Stewart Blusson Quantum Matter Institute, University of British Columbia,
Vancouver, British Columbia V6T 1Z4, Canada }

\date{\today}
\begin{abstract}
We revisit the importance of collective-mode fluctuations and gauge
invariance in the electromagnetic response of superconducting systems.
In particular, we show that order-parameter fluctuations, gapless
or not, have no contribution to the Meissner effect in both $s$-
and $p$-wave superconductors. More generally, we extend this result
to uniform and nonuniform superfluids with no external wavevector
scale. To facilitate this analysis, we formulate a path-integral-based
matrix methodology for computing the electromagnetic response of fermionic
fluids in the presence of concomitantly fluctuating collective modes.
Closed-form expressions for the electromagnetic response in different
scenarios are provided, including the case of fluctuations of electronic
density and the phase and amplitude of the order parameter. All microscopic
symmetries and invariances are manifestly satisfied in our formalism,
and it can be straightforwardly extended to other scenarios. 
\end{abstract}
\maketitle

\section{Introduction \label{sec:Introduction}}

Collective modes in superfluids and superconductors play a pivotal
role in understanding gauge invariance in a many-particle context~\citep{Schrieffer,Rickayzen,Parks}.
These modes comprise amplitude and phase fluctuations of the order
parameter~\citep{Arseev_2006,Anderson_2016}, and in the context
of neutral superfluids the presence of the phase mode is evinced as
a longitudinal sound oscillation~\citep{Anderson_1958,Rickayzen_1959,Nambu_1960}.
Observation of the amplitude mode in a condensed-matter context, while
possible, is rather challenging~\citep{Pekker_Varma_2015}. Some
particular cases where this mode was indeed observed include systems
with emergent Lorentz invariance~\citep{Pollet_Prokofev_2012,Bloch_2012,Sherman_2015}
and superconductors coupled to either charge-density waves~\citep{Littlewood_Varma_1981,Browne_Levin_1983}
or optical modes~\citep{Matsunaga1145,shimano_higgs}. Collective
modes in general provide non-trivial examples of the rich physics
associated with broken symmetries and non-trivial ordering~\citep{RevModPhys.87.457,Zha_1995}.

In contrast, the Meissner effect is conventionally understood as a
``transverse'' response~\citep{PinesNozieres2,Martin_1967}, where
``longitudinal'' collective modes are not thought to participate.
This issue was addressed, partly clarified, in Ref.~\citep{Millis_1987}.
There it was shown that in nonuniform superfluids the ``longitudinal'' collective 
modes can possibly appear in what are termed -- in the context of 
uniform systems -- ``transverse'' response functions. In addition, Ref.~\citep{Boyack_2017} 
provided an explicit calculation of the electromagnetic (EM) response of the
Fulde-Ferrell (FF) superfluid, which consists of finite-momentum Cooper
pairs, and showed that the amplitude mode gives a significant contribution
to the superfluid density. These issues motivate the current work,
where we investigate the superfluid response for systems with nonuniform
pairing, such as $p$-wave superfluids~\citep{Hoyos_Moroz_Son_2014,Ariad_2015}
and superconductors~\citep{Yakovenko_2008,Yakoveko_2009}, and we
provide a more general understanding on the type of superconductor
where collective modes can contribute to the Meissner response. 

We define a uniform superfluid or superconductor to be one where the
order parameter two-point function is both translation and rotation
invariant. A nonuniform system is one that is not uniform, and as
such it violates either one or both of the conditions above. In the
case of uniform $s$-wave superconductors, gauge invariance and the
uniformity of the gap establishes that there is no collective-mode
contribution to the Meissner effect~\citep{AGDBook}. When isotropy
is broken, however, this argument needs to be revisited~\citep{Millis_1987}.

\textcolor{black}{Phase fluctuations of the order parameter must be
included to derive a gauge-invariant EM response~\citep{Anderson_2016}.
On top of this, one can also consider amplitude fluctuations of the
order parameter, and these have been shown~\citep{Anderson_2016}
to be necessary to satisfy a thermodynamic sum rule, namely the compressibility
sum rule~\citep{Guo_Chien_He_2013,PinesNozieres1}. }Of particular
interest is the response in $p$-wave superfluids~\citep{Ariad_2015,Hoyos_Moroz_Son_2014,Yakovenko_2008,Yakoveko_2009}
and also in systems with other pairing symmetries~\citep{Sharapov_2001}.
A complete calculation of the EM response for a $p$-wave system,
in the presence of Coulomb and amplitude and phase\textcolor{black}{{}
fluctuations of the order parameter, has not been, to the best of
our knowledge, presented in the literature, and the question of the
Meissner response for such a system was unaddressed in Ref.~\citep{Millis_1987}.
In this paper we show that collective modes do not contribute to the
Meissner effect in neither} uniform\textcolor{black}{{} }$s$-\textcolor{black}{wave
nor nonuniform }$p$-wave \textcolor{black}{superconductors}. More
generally, our results show that collective modes do not contribute
to the Meissner effect, independent of the pairing symmetry, in any
superconductor that does not display an external wavevector scale
(e.g. finite-momentum pairing).

In order to derive this result, we develop a method for computing
the gauge-invariant EM response of an electronic system with multiple
collective modes present. Our analysis is based on an extension of
the path-integral formulation of Ref.~\citep{Anderson_2016} and
matrix linear-response approaches of Refs.~\citep{Kulik1981,Arseev_2006,Guo_Chien_He_2013}.
One of our central results is to demonstrate how these collective
modes can be incorporated in comprehensive and illuminating EM response
tensors using singular-value decompositions. For pedagogical purposes
we consider several examples of application, including the Coulomb
screening in a normal metal and phase fluctuations in the EM response
of a superfluid. We demonstrate the power of our formulation by obtaining
the manifestly gauge-invariant EM response tensor for superconductors
with amplitude, phase, and Coulomb fluctuations present. More generally,
our results are applicable to a variety of scenarios beyond the scope
of this work. They are relevant in any situation where energy scales
compete, leading to intertwined ordering~\citep{RevModPhys.87.457},
or where symmetries provide multi-dimensional order parameters. The
study of the concomitant contribution of distinct collective modes
to the EM response tensor provides a direct method to access signatures
of broken symmetries and non-trivial ordering.

The paper is organized as follows: in Sec.~\ref{sec:Responses} we
outline general formulae for the electromagnetic susceptibility tensors;
a careful derivation of these formulae is provided in Appendix~\ref{sec:tensor}.
Following this, Sec.~\ref{sec:Applications} provides a set of applications
of these formulae, including: Coulomb screening, phase fluctuations
in a superfluid, the gapping of phase modes in a superconductor by
Coulomb screening, and finally the mixing of phase modes with amplitude-Higgs
modes in a charged superconductor. This section contains our algebraic
approach to screening by use of singular-value decompositions. Finally,
Sec.~\ref{sec:Meissner} addresses our discussions regarding the
Meissner effect and we conclude in Sec.~\ref{sec:Conclusions}. Appendices~\ref{sec:Qmunu}-\ref{sec:Det_SVD}
provide further details on several relevant calculations.

\section{Electromagnetic response tensor \label{sec:Responses}}

The starting point of our analysis is a fermionic system subject to
a set of collective fluctuating degrees of freedom. The latter are
described by a set of generalized coordinates, denoted by $\boldsymbol{\Delta}$,
which should be thought of as a vector of Hubbard-Stratonovich decoupling
fields. In the presence of an external EM probe $A$, we consider
the dynamics of the EM response at the mean-field level, which is
defined by the following conditions for each component $\Delta_{a}$
of $\boldsymbol{\Delta}$ :
\begin{equation}
\left.\frac{\delta S_{\text{eff}}\left[\boldsymbol{{\bf \Delta}},A\right]}{\delta{\bf \Delta}_{a}\left(x\right)}\right|_{\boldsymbol{\Delta}=\boldsymbol{{\bf \Delta}}_{\textrm{mf}}\left[A\right]}=0.\label{eq:MF}
\end{equation}
Here, $S_{\text{eff}}$ is the effective action for the fluctuating
degrees of freedom, in the presence of the external EM probe, obtained
after integration over the fermionic degrees of freedom~\footnote{We assume that the procedure of integrating out the fermions is well
defined.}. The solutions to the mean-field equations, $\boldsymbol{\Delta}_{\text{mf}}\left[A\right]$,
are no longer arbitrary fluctuating degrees of freedom to be functionally
integrated over, but rather they are functions determined by the external
EM probe~\citep{Ambegaokar_Kadanoff_1961,Arseev_2006,Anderson_2016}.
As a result, the mean-field EM response tensor reads 

\begin{equation}
K_{\textrm{mf}}^{\mu\nu}(x,y)=\left.\frac{\delta^{2}S_{\text{eff}}\left[\boldsymbol{{\bf \Delta}}_{\textrm{mf}}\left[A\right],A\right]}{\delta A_{\mu}\left(x\right)\delta A_{\nu}\left(y\right)}\right|_{A=0}.
\end{equation}
Note that $K^{\mu\nu}\left(x,y\right)=K^{\nu\mu}\left(y,x\right)$.
In this paper imaginary time will be used and thus $A^{\mu}=\left(A_{0},\mathbf{A}\right)=\left(iA_{t},\mathbf{A}\right)$. 

To evaluate these derivatives it is necessary to use a functional
chain rule and differentiate all terms with dependence on the vector
potential. This manipulation, together with an application of the
mean-field equations in Eq.~(\ref{eq:MF}), is presented in Appendix~\ref{sec:tensor};
the result is a matrix form for the mean-field-level EM response,
namely,

\begin{align}
K_{\textrm{mf}}^{\mu\nu}\left(x,y\right)= & Q^{\mu\nu}\left(x,y\right)-\int_{z,z'}\biggl\{ R^{\mu a}\left(x,z\right)\nonumber \\
 & \times\left[S^{-1}\left(z,z'\right)\right]^{ab}R^{b\nu}\left(z',y\right)\biggr\},\label{eq:EM_resp}
\end{align}

where

\begin{align}
Q^{\mu\nu}\left(x,y\right) & =\left.\frac{\delta^{2}S_{\text{eff}}\left[\boldsymbol{{\bf \Delta}},A\right]}{\delta A_{\mu}\left(x\right)\delta A_{\nu}\left(y\right)}\right|_{A=0,\boldsymbol{{\bf \Delta}}=\boldsymbol{{\bf \Delta}}_{\textrm{mf}}[0]},\label{eq:Qdef}\\
R^{\mu a}\left(x,y\right) & =R^{a\mu}\left(y,x\right)\nonumber \\
 & =\left.\frac{\delta^{2}S_{\text{eff}}\left[\boldsymbol{{\bf \Delta}},A\right]}{\delta A_{\mu}\left(x\right)\delta{\bf \Delta}_{a}\left(y\right)}\right|_{A=0,\boldsymbol{{\bf \Delta}}=\boldsymbol{{\bf \Delta}}_{\textrm{mf}}[0]},\label{eq:Rdef}
\end{align}
and
\begin{equation}
S^{ab}\left(x,y\right)=\left.\frac{\delta^{2}S_{\text{eff}}\left[\boldsymbol{{\bf \Delta}},A\right]}{\delta{\bf \Delta}_{a}\left(x\right)\delta{\bf \Delta}_{b}\left(y\right)}\right|_{A=0,\boldsymbol{{\bf \Delta}}=\boldsymbol{{\bf \Delta}}_{\textrm{mf}}[0]}.\label{eq:Sdef}
\end{equation}
Here, the derivatives with respective to the gauge field $A$ act
only on the explicit vector-potential dependence. In the second contribution
of Eq.~(\ref{eq:EM_resp}), we emphasize that the matrix $S^{ab}$
must be computed first, as in Eq.~(\ref{eq:Sdef}), and then inverted
before being inserted into Eq.~(\ref{eq:EM_resp}). In other words,
Eq.~(\ref{eq:EM_resp}) does not involve the inverse of each matrix
element of Eq.~(\ref{eq:Sdef}), but rather the elements of the inverse
of the matrix itself.

This expression contains several insightful properties. First, it
manifestly decouples into two contributions which correspond, respectively,
to the bubble and collective-mode linear responses. Second, this expression
is reparameterization covariant, i.e., it does not change form under
a basis transformation of $\boldsymbol{\Delta}$. This means that
all fluctuations are considered symmetrically, in an unbiased manner.
In the context of superconductivity, for example, Eq.~(\ref{eq:EM_resp})
can be equally used for considering fluctuations in the real and imaginary
part of the superconducting pairing strength~\citep{Kulik1981},
or for fluctuations in the radial and phase degrees of freedom, as
we shall do later in the paper. Third, by writing this expression
in real space it affords greater generality and can thus be used,
for example, in the presence of either impurities or defects occurring
in collective-mode order parameters. For a translation-invariant system,
the momentum-space representation is more tractable and reads
\begin{equation}
K_{\text{mf}}^{\mu\nu}\left(q\right)=Q^{\mu\nu}\left(q\right)-R^{\mu a}\left(q\right)\left[S^{-1}\left(q\right)\right]^{ab}R^{b\nu}\left(q\right),\label{eq:Em_in_q}
\end{equation}
where, for example, 

\begin{equation}
Q^{\mu\nu}\left(x,y\right)=Q^{\mu\nu}\left(x-y\right)=\int_{q}e^{-iq\cdot\left(x-y\right)}Q^{\mu\nu}\left(q\right).
\end{equation}
We use the short-hand notation $\int_{q}=TL^{d}\sum_{i\Omega_{m}}\int\frac{d{\bf q}}{\left(2\pi\right)^{d}}$,
where $L$ is a length scale, $d$ is the number of spatial dimensions,
$T$ is the temperature, and $\Omega_{m}$ is a bosonic Matsubara
frequency. Natural units $c=\hbar=k_{B}=1$ are used throughout the
paper. 

\section{General applications \label{sec:Applications}}

In this section we present several applications of Eq.~(\ref{eq:Em_in_q}).
For the benefit of the reader, in the following subsections we take
a pedagogical approach and start with a rather detailed calculation
of the application of Eq.~(\ref{eq:EM_resp}) in two familiar scenarios:
\ref{subsec:ScreeningCoulomb}-Electrostatic screening and \ref{subsec:SCPhase}-gauge-invariant
response in superfluids due to phase fluctuations. With the mathematical
procedures well established, we will then move on at a progressively
faster pace: in \ref{subsec:ChargedSCPhase} we study the next simplest
possible scenario -- a superconductor with phase fluctuations --
and here we introduce the concept of folding the effects of competing
fluctuations using singular-value decompositions. The d$\acute{\text{e}}$nouement
of this section is \ref{subsec:ChargedSCPhaseAmplitude}, where we
put all this methodology together to compute the EM response tensor
in the non-trivial case of concomitantly fluctuating Coulomb and superconducting
phase and amplitude degrees of freedom. To clarify our terminology,
a superconductor is a charged system with Coulomb interactions present
and a superfluid is a neutral system. 

\subsection{Screening due to electrostatic interactions\label{subsec:ScreeningCoulomb}}

Consider an interacting electronic system in $D=d+1$ spacetime dimensions
with an action given by
\begin{align}
S\left[A\right]= & -\int d^{D}xd^{D}x'\psi_{\sigma}^{\dagger}\left(x\right)\mathcal{G}_{0}^{-1}\left[A\right]\left(x,x'\right)\psi_{\sigma}\left(x'\right)\nonumber \\
 & +\frac{e^{2}}{2}\int d^{D}xd^{D}x'\delta n\left(x\right)V\left(x-x'\right)\delta n\left(x'\right)\nonumber \\
 & +ie\int d^{D}xA_{t}\left(x\right)n_{0},
\end{align}
where $\delta n\left(x\right)=\psi_{\sigma}^{\dagger}\left(x\right)\psi_{\sigma}\left(x\right)-n_{0}$,
with $n_{0}$ the constant background density, $\sigma=\downarrow,\uparrow$
is a spin index (summed if repeated) and the inverse Green's function
is

\begin{equation}
\mathcal{G}_{0}^{-1}\left[A\right]\left(x,x'\right)=-\left(\partial_{\tau}-ieA_{t}\left(x\right)+h\left(\hat{\mathbf{p}}-e\mathbf{A}\right)\right)\delta\left(x-x'\right).
\end{equation}
The single-particle Hamiltonian, denoted by $h\left({\bf p}\right)$,
is kept general at this stage. For concreteness, we assume instantaneous
interactions: $V\left(x-x'\right)=V\left(\mathbf{x}-\mathbf{x}'\right)\delta\left(\tau-\tau'\right)$.
Throughout the paper we shall interchangeably refer to electronic
density fluctuations as Coulomb fluctuations. The generating functional
for electromagnetic response is then
\begin{equation}
\mathcal{Z}\left[A\right]=\int\mathcal{D}\left[\psi^{\dagger},\psi\right]e^{-S\left[A\right]}.
\end{equation}

We are interested in how nonuniform charge distributions affect the
EM response of this system. Thus it is natural to consider decoupling
the electrostatic interaction terms via a Hubbard-Stratonovich decomposition
as 

\begin{equation}
\mathcal{Z}\left[A\right]\sim\int\mathcal{D}\varphi e^{-S_{\text{eff}}\left[\varphi,A\right]},
\end{equation}
Defining $\beta=1/T$ and $\mathcal{G}^{-1}\left[\varphi,A\right]=\mathcal{G}^{-1}\left[A_{t}+\varphi,\mathbf{A}\right]$,
the effective action is 

\begin{align}
S_{\text{eff}}\left[\varphi,A\right]= & \int d^{D}xd^{D-1}x'\frac{\varphi\left(\mathbf{x},\tau\right)\varphi\left(\mathbf{x}',\tau\right)}{2V\left(\mathbf{x}-\mathbf{x}'\right)}\nonumber \\
 & +ie\int d^{D}x\left(A_{t}\left(x\right)+\varphi\left(x\right)\right)n_{0}\nonumber \\
 & -\textrm{Tr}\ln\left(-\beta\mathcal{G}_{0}^{-1}\left[\varphi,A\right]\right).
\end{align}
The capitalized trace denotes a trace over all space-time/momentum-frequency
and internal (uncapitalized trace) degrees of freedom:
\begin{equation}
\textrm{Tr}\ln\left(-\beta\mathcal{G}_{0}^{-1}\left[\varphi,A\right]\right)=\int d^{D}x\text{tr}\left\langle x\left|\ln\left(-\beta\mathcal{G}_{0}^{-1}\left[\varphi,A\right]\right)\right|x\right\rangle .\label{eq:trace_eff}
\end{equation}
In this language, we obtain the building blocks for Eq.~(\ref{eq:EM_resp})
(which are tantamount to undressed polarization tensors). In fact,
due to translation invariance, we can focus on the expressions in
momentum space used in Eq.~(\ref{eq:Em_in_q}). For instance,

\begin{align}
Q^{\mu\nu}\left(q\right) & \equiv\left.\frac{\delta^{2}S_{\text{eff}}\left[\varphi,A\right]}{\delta A_{\mu}\left(-q\right)\delta A_{\nu}\left(q\right)}\right|_{A,\varphi=0}\nonumber \\
 & =-\left.\frac{\delta^{2}\textrm{Tr}\ln\left(-\beta\mathcal{G}_{0}^{-1}\left[\varphi,A\right]\right)}{\delta A_{\mu}\left(-q\right)\delta A_{\mu}\left(q\right)}\right|_{A,\varphi=0}.
\end{align}
Similarly, noticing that $A_{0}=iA_{t}$ and that all terms involving
$\varphi$ appear in the Green's function as $iA_{t}+i\varphi$ ,
one finds
\begin{align}
R^{\mu\varphi}\left(q\right) & \equiv\left.\frac{\delta^{2}S_{\text{eff}}\left[\varphi,A\right]}{\delta A_{\mu}\left(-q\right)\delta\varphi\left(q\right)}\right|_{A,\varphi=0}=iQ^{\mu0}\left(q\right),\\
S^{\varphi\varphi}\left(q\right) & =\left.\frac{\delta^{2}S_{\text{eff}}\left[\boldsymbol{{\bf \Delta}},A\right]}{\delta\varphi\left(-q\right)\delta\varphi\left(q\right)}\right|_{A,\varphi=0}\nonumber \\
 & =V^{-1}\left(q\right)-Q^{00}\left(q\right).\label{eq:Sfifi}
\end{align}
Conveniently, all building blocks can be expressed in terms of the
undressed polarization tensor $Q^{\mu\nu}\left(q\right)$. An in-depth
analysis of these expressions is provided in Appendix~\ref{sec:Qmunu}.

Applying Eq.~(\ref{eq:Em_in_q}) now becomes a simple matter (we
drop the $q$-dependence label for simplicity):

\begin{align}
K_{\text{mf}}^{\mu\nu} & =Q^{\mu\nu}-\left(iQ^{\mu0}\right)\left(V^{-1}-Q^{00}\right)^{-1}\left(iQ^{0\nu}\right)\nonumber \\
 & =Q^{\mu\nu}+\frac{Q^{\mu0}VQ^{0\nu}}{1-VQ^{00}}\equiv\widetilde{Q}^{\mu\nu}.\label{eq:Coulomb_scre}
\end{align}
The last definition will be used throughout later sections of the
paper. The above result reproduces the screening effect of Coulomb
fluctuations. In particular, the RPA charge-charge susceptibility~\citep{PinesNozieres1}
is obtained:
\begin{equation}
K_{\text{mf}}^{00}=\frac{Q^{00}}{1-VQ^{00}}.
\end{equation}

\subsection{EM response for superfluids (with no amplitude fluctuations)\label{subsec:SCPhase}}

Another simple application of Eq.~(\ref{eq:Em_in_q}) concerns the
gauge-invariant EM response tensor for superfluids with phase fluctuations
of the order parameter. In superfluids where the mean-field order
parameter takes on a finite vacuum expectation value the global U(1)
symmetry is spontaneously broken. To restore gauge-invariance, the
phase fluctuations of the order parameter must be included. In this
section we consider a superfluid where the amplitude of the order
parameter is rigidly pinned down to its mean-field value, but allow
the phase to depend on the external EM probe. 

It is straightforward to analyze this scenario with our present approach.
Consider a set of non-relativistic spin-$\frac{1}{2}$ particles,
with free Hamiltonian $h\left(\mathbf{p}\right)=\mathbf{p}^{2}/\left(2m\right)-\mu$,
interacting instantaneously with each other via an attractive, translation-invariant,
but possibly anisotropic potential $g\left(\mathbf{x}-\mathbf{x}^{\prime}\right)$.
In the presence of an external probe field $A$, the action reads 

\begin{align}
S\left[A\right]= & -\int d^{D}xd^{D}x'\psi_{\sigma}^{\dagger}\left(x\right)\mathcal{G}_{0}^{-1}\left[A\right]\left(x,x'\right)\psi_{\sigma}\left(x'\right)\nonumber \\
 & -\int d^{D}xd^{D}x'\psi_{\uparrow}^{\dagger}\left(x\right)\psi_{\downarrow}^{\dagger}\left(x'\right)g\left(x-x^{\prime}\right)\psi_{\downarrow}\left(x'\right)\psi_{\uparrow}\left(x\right)\nonumber \\
 & +ie\int d^{D}xA_{t}n_{0}.
\end{align}
Here, $g\left(x-x^{\prime}\right)=g\left(\mathbf{x}-\mathbf{x}^{\prime}\right)\delta\left(\tau-\tau^{\prime}\right)$.

Preparing again for the mean-field treatment of the problem, we now
perform a Hubbard-Stratonovich decomposition in the Cooper channel
to arrive at the generating functional

\begin{equation}
\mathcal{Z}\left[A\right]\sim\int\mathcal{D}\left[\Delta,\Delta^{*}\right]\mathcal{D}\left[\psi^{\dagger},\psi\right]e^{-S_{\text{bos}}}e^{-S_{\text{el}}},
\end{equation}
where the bosonic contribution to the action is 
\begin{equation}
S_{\text{bos}}=ie\int d^{D}xA_{t}n_{0}+\int d^{D}xd^{D-1}x^{\prime}\frac{\left|\Delta\left(\mathbf{x},\mathbf{x}^{\prime},\tau\right)\right|^{2}}{g\left(\mathbf{x}-\mathbf{x}^{\prime}\right)}
\end{equation}
and the electronic contribution is
\begin{align}
S_{\text{el}}= & -\int d^{D}xd^{D}x^{\prime}\psi_{\sigma}^{\dagger}\left(x\right)\mathcal{G}_{0}^{-1}\left[A\right]\left(x,x^{\prime}\right)\psi_{\sigma}\left(x^{\prime}\right)\nonumber \\
 & -\int d^{D}xd^{D-1}x^{\prime}\left[\psi_{\uparrow}^{\dagger}\left(\mathbf{x},\tau\right)\Delta\left(\mathbf{x},\mathbf{x}^{\prime},\tau\right)\psi_{\downarrow}^{\dagger}\left(\mathbf{x}^{\prime},\tau\right)+\text{h.c.}\right].
\end{align}
Before integrating out the fermions, remember that the symmetry of
the interaction potential $g\left(\mathbf{x}-\mathbf{x}^{\prime}\right)$
is decisive in determining the symmetry structure of the pairing field.
Due to the homogeneity of the problem (in the absence of strong driving
external EM fields), it is advantageous to use relative and center-of-mass
coordinates to describe the pairing field:
\begin{equation}
\Delta\left(\mathbf{x},\mathbf{x}^{\prime},\tau\right)\to\Delta\left(\mathbf{x}-\mathbf{x}^{\prime},\frac{\mathbf{x}+\mathbf{x}^{\prime}}{2},\tau\right).
\end{equation}
We ignore spin-orbit coupling. In this case, spherical anisotropy
in the pairing potential can be captured in a gradient expansion of
$\Delta$

\begin{align}
 & \Delta\left(\mathbf{x}-\mathbf{x}^{\prime},\frac{\mathbf{x}+\mathbf{x}^{\prime}}{2},\tau\right)\nonumber \\
 & =\left|\Delta_{s}\left(\frac{\mathbf{x}+\mathbf{x}^{\prime}}{2},\tau\right)\right|e^{i\Phi_{s}\left(\frac{\mathbf{x}+\mathbf{x}^{\prime}}{2},\tau\right)}\delta\left(\mathbf{x}-\mathbf{x}^{\prime}\right)\label{eq:pairing_exp}\\
 & +\left|\Delta_{p}\left(\frac{\mathbf{x}+\mathbf{x}^{\prime}}{2},\tau\right)\right|e^{i\Phi_{p}\left(\frac{\mathbf{x}+\mathbf{x}^{\prime}}{2},\tau\right)}\left(\partial_{x}+i\partial_{y}\right)\delta\left(\mathbf{x}-\mathbf{x}^{\prime}\right)+...,\nonumber 
\end{align}
where we favor an amplitude-phase coordinate choice. In general, the
pairing potential will select only one term in Eq.~(\ref{eq:pairing_exp});
the structure we chose for the interaction, in fact, favors opposite-spin
pairing by construction. Nevertheless, we can remain fairly general
and write

\begin{align}
S_{\text{bos}} & =ie\int d^{D}xA_{t}n_{0}+\int d^{D}x\frac{\left|\Delta\left(\mathbf{x},\tau\right)\right|^{2}}{2\widetilde{g}},\label{eq:SC_bos_act}
\end{align}
where $\widetilde{g}$ is a renormalized value for $g$, and
\begin{align}
S_{\text{el}}= & -\int d^{D}xd^{D}x^{\prime}\psi_{\sigma}^{\dagger}\left(x\right)\mathcal{G}_{0}^{-1}\left[A\right]\left(x,x^{\prime}\right)\psi_{\sigma}\left(x^{\prime}\right)\nonumber \\
 & -\int d^{D}x\left[\Delta\left(\mathbf{x},\tau\right)\psi_{\uparrow}^{\dagger}\left(\mathbf{x},\tau\right)\hat{D}\psi_{\downarrow}^{\dagger}\left(\mathbf{x},\tau\right)+\text{h.c.}\right],\label{eq:SC_exp_act}
\end{align}
where $\Delta\left(\mathbf{x},\tau\right)=\rho\left(x\right)e^{i\theta\left(x\right)}$
for a general amplitude and phase and $\hat{D}$ corresponds to a
differential operator that depends on the symmetry channel. In Appendix~\ref{sec:Pairing_app}
we consider an explicit application of this to a spinless $p$-wave
problem.

We are now ready to integrate out the fermions; introducing a Nambu
doubled spinor $\Psi=\left(\psi_{\uparrow},\psi_{\downarrow},\psi_{\uparrow}^{\dagger},\psi_{\downarrow}^{\dagger}\right)^{T}$,
the electronic part of the action becomes

\begin{equation}
S_{\text{el}}=-\frac{1}{2}\int d^{D}xd^{D}x'\Psi^{\dagger}\left(x\right)\mathcal{G}^{-1}\left[A\right]\left(x,x'\right)\Psi\left(x'\right)
\end{equation}
where the (inverse) Nambu-space Green's function is

\begin{widetext}

\begin{align}
\mathcal{G}^{-1}\left[A\right]\left(x,x'\right) & =-\left(\begin{array}{cc}
\left[\partial_{\tau}-ie\widetilde{A}_{t}+\left(\frac{\left[\hat{\mathbf{p}}-e\widetilde{\mathbf{A}}\right]^{2}}{2m}-\mu\right)\right] & -\rho\left(x\right)i\sigma_{y}\hat{D}\\
\rho\left(x\right)i\sigma_{y}\hat{D}^{\dagger} & \left[\partial_{\tau}+ie\widetilde{A}_{t}-\left(\frac{\left[\hat{\mathbf{p}}+e\widetilde{\mathbf{A}}\right]^{2}}{2m}-\mu\right)\right]
\end{array}\right)\delta\left(x-x'\right),\label{eq:SC_GF}
\end{align}

\end{widetext} $\sigma_{y}$ acts on the spin degrees of freedom,
and we have rotated away the superconducting phase, which is conveniently
absorbed by the gauge fields as $\widetilde{A}_{\mu}=A_{\mu}-\frac{1}{2e}\partial_{\mu}\theta$.
The generating functional thus becomes
\begin{equation}
\mathcal{Z}\left[A\right]\sim\int\mathcal{D}\left[\Delta,\Delta^{*}\right]e^{-S_{\mathrm{eff}}\left[\Delta,\Delta^{*},A\right]},
\end{equation}
where the effective action is (dropping the $A$-dependence label)

\begin{eqnarray}
S_{\text{eff}}\left[\Delta,\Delta^{*},A\right] & = & S_{\text{bos}}-\frac{1}{2}\textrm{Tr}\ln\left(-\beta\mathcal{G}^{-1}\right),
\end{eqnarray}
with $S_{\mathrm{bos}}$ as in Eq.~(\ref{eq:SC_bos_act}) and one
should keep in mind the factor of $\frac{1}{2}$ due to Nambu doubling. 

At this point we consider the mean-field response. In this section,
we will neglect fluctuations of the superconducting amplitude, setting
$\rho\left(x\right)\to\rho_{0}$. It is then possible to use the relationship
between $\widetilde{A}_{\mu}$ and $A_{\mu}$ to write

\begin{eqnarray}
\frac{\delta S_{\text{eff}}\left[\theta,A\right]}{\delta\theta\left(x\right)} & = & \int dy\frac{\delta S_{\text{eff}}\left[\theta,A\right]}{\delta\partial_{\alpha}\theta\left(y\right)}\frac{\delta\partial_{\alpha}\theta\left(y\right)}{\delta\theta\left(x\right)}\nonumber \\
 & = & -\partial_{\alpha}\frac{\delta S_{\text{eff}}\left[\theta,A\right]}{\delta\partial_{\alpha}\theta\left(x\right)}\nonumber \\
 & = & \frac{1}{2e}\partial_{\alpha}\frac{\delta S_{\text{eff}}\left[\theta,A\right]}{\delta A_{\alpha}\left(x\right)}.
\end{eqnarray}
The factor of $2e$ can be safely absorbed as it will drop out from
the correlation functions; we will omit it from now on. This allows
us to once again write all the momentum-space tensors in terms of
the undressed polarization tensors $Q^{\mu\nu}$, namely,
\begin{align}
R^{\mu\theta}\left(q\right) & =iQ^{\mu\beta}\left(q\right)q_{\beta},\\
R^{\theta\nu}\left(q\right) & =-iq_{\alpha}Q^{\alpha\nu}\left(q\right),\\
S^{\theta\theta}\left(q\right) & =q_{\lambda}Q^{\lambda\sigma}\left(q\right)q_{\sigma}.
\end{align}
At the mean-field level $\theta$ is a constant and drops out from
the Green's functions. Notice that the Green's functions appearing
in $Q^{\mu\nu}\left(q\right)$ in this case correspond to Eq.~(\ref{eq:SC_GF})
with $\widetilde{A}_{\mu}=0$ and $\rho\left(x\right)\to\rho_{0}$.
Implementing Eq.~(\ref{eq:Em_in_q}), the EM response is then

\begin{align}
K_{\text{mf}}^{\mu\nu} & =Q^{\mu\nu}-\left(iQ^{\mu\beta}q_{\beta}\right)\left(q_{\lambda}Q^{\lambda\sigma}q_{\sigma}\right)^{-1}\left(-iq_{\alpha}Q^{\alpha\nu}\right)\nonumber \\
 & =Q^{\mu\nu}-\frac{Q^{\mu\beta}q_{\beta}q_{\alpha}Q^{\alpha\nu}}{q_{\lambda}Q^{\lambda\sigma}q_{\sigma}}\equiv\Pi^{\mu\nu}.\label{eq:phase_scre}
\end{align}
This is the general form of the EM response tensor for a neutral superfluid,
independent of the pairing symmetry. The gapless fluctuating phase
degree of freedom is crucial to ensure gauge invariance, which the
form above manifestly obeys: $q_{\mu}K_{\text{mf}}^{\mu\nu}\left(q\right)=K_{\text{mf}}^{\mu\nu}\left(q\right)q_{\nu}=0$.
Setting $q_{\lambda}Q^{\lambda\sigma}q_{\sigma}=0$ recovers the well-known
result of Anderson and Bogoliubov~\citep{Anderson_1958b,Bogoliubov1958,*BogoliubovBook}:
the EM response has a pole corresponding to a long-wavelength sound
mode (with speed $c_{s}=v_{F}/\sqrt{3}$ at $T=0$) induced by phase
fluctuations of the order parameter. 

\subsection{EM response for superconductors (with no amplitude fluctuations)\label{subsec:ChargedSCPhase}}

With the previous results established, for our first non-trivial application
of Eq.~(\ref{eq:Em_in_q}) we consider a charged superconductor with
both phase and Coulomb fluctuations present. This problem was also
considered in Ref.~\citep{Yakovenko_2008}, in the context of the
EM response of a $p$-wave superconductor, via sequential functional
integration of the Coulomb and phase degrees of freedom. It is natural
to ask what the form of the EM response would be if this procedure
were performed in the opposite order, and this will be addressed in
what follows. In our case, the results from the previous sections
allow the response to be written as
\begin{align}
K_{\text{mf}}^{\mu\nu}= & Q^{\mu\nu}-\left(\begin{array}{c}
iQ^{\mu0}\\
iq_{\beta}Q^{\mu\beta}
\end{array}\right)^{T}\nonumber \\
 & \times\left(\begin{array}{cc}
V^{-1}-Q^{00} & -Q^{0\beta}q_{\beta}\\
q_{\alpha}Q^{\alpha0} & q_{\lambda}Q^{\lambda\sigma}q_{\sigma}
\end{array}\right)^{-1}\left(\begin{array}{c}
iQ^{0\nu}\\
-iq_{\alpha}Q^{\alpha\nu}
\end{array}\right)\nonumber \\
= & Q^{\mu\nu}-\frac{1}{\left(V^{-1}-Q^{00}\right)q_{\lambda}\widetilde{Q}^{\lambda\sigma}q_{\sigma}}\left(\begin{array}{c}
Q^{\mu0}\\
q_{\beta}Q^{\mu\beta}
\end{array}\right)^{T}\nonumber \\
 & \times\left(\begin{array}{cc}
-q_{\lambda}Q^{\lambda\sigma}q_{\sigma} & q_{\beta}Q^{0\beta}\\
q_{\alpha}Q^{\alpha0} & V^{-1}-Q^{00}
\end{array}\right)\left(\begin{array}{c}
Q^{0\nu}\\
q_{\alpha}Q^{\alpha\nu}
\end{array}\right).\label{eq:AB1}
\end{align}
The Coulomb-screened EM response tensor $\widetilde{Q}^{\lambda\sigma}$
of Sec.~\ref{subsec:ScreeningCoulomb} naturally appears here in
the denominator.

While Eq.~(\ref{eq:AB1}) treats the Coulomb- and phase-screened
responses of a charged superconductor in a symmetric fashion, the
present form is not totally satisfactory. In particular, gauge invariance
is not manifest, and it may be advantageous to recover similar results
found in the previous section, as well as the polaritonic resonances
of the EM response. To accomplish this, we have to ``bias'' the
above expression towards either a Coulomb-screened type of object
or a phase-screened type of object. An analogy from the process of
Ref.~\citep{Yakovenko_2008} would be to consider integrating out
first either the electrostatic Coulomb field or the phase degree of
freedom. 

Let us make this procedure more explicit. With a few manipulations,
we may explicitly rewrite $Q^{\mu\nu}$ in terms of its Coulomb-screened
version $\widetilde{Q}^{\mu\nu}$ so that Eq.~(\ref{eq:AB1}) then
has the form
\begin{align}
K_{\text{mf}}^{\mu\nu}= & \widetilde{Q}^{\mu\nu}-\frac{q_{\alpha}q_{\beta}}{q_{\lambda}\widetilde{Q}^{\lambda\sigma}q_{\sigma}}\nonumber \\
 & \times\left(\begin{array}{c}
Q^{\mu0}\\
Q^{\mu\beta}
\end{array}\right)^{T}\left(\begin{array}{cc}
\frac{Q^{\alpha0}Q^{0\beta}}{\left(V^{-1}-Q^{00}\right)^{2}} & \frac{Q^{0\beta}}{V^{-1}-Q^{00}}\\
\frac{Q^{\alpha0}}{V^{-1}-Q^{00}} & 1
\end{array}\right)\left(\begin{array}{c}
Q^{0\nu}\\
Q^{\alpha\nu}
\end{array}\right).\label{eq:Coulomb_biasing}
\end{align}
The $2\times2$ matrix appearing in the EM response now has zero
determinant: it is a singular matrix, which can be expressed using
a singular-value decomposition (SVD). Consider the following matrix
\begin{equation}
M=\left(\begin{array}{cc}
ab & a\\
b & 1
\end{array}\right).
\end{equation}
Define the matrices $U,V$, and $D$ by
\begin{equation}
U=\left(\begin{array}{cc}
a & \frac{a}{\left|a\right|}\\
1 & -\left|a\right|
\end{array}\right),V=\left(\begin{array}{cc}
b^{*} & \frac{b^{*}}{\left|b\right|}\\
1 & -\left|b\right|
\end{array}\right),D=\left(\begin{array}{cc}
1 & 0\\
0 & 0
\end{array}\right).
\end{equation}
The matrix $M$ can then be written as $M=UDV^{\dagger}.$ By matching
the coefficients $a$ and $b$ with the coefficients in Eq.~(\ref{eq:Coulomb_biasing}),
one obtains

\begin{equation}
K_{\mathrm{mf}}^{\mu\nu}=\widetilde{Q}^{\mu\nu}-\frac{\widetilde{Q}^{\mu\beta}q_{\beta}q_{\alpha}\widetilde{Q}^{\alpha\nu}}{q_{\lambda}\widetilde{Q}^{\lambda\sigma}q_{\sigma}}\equiv\widetilde{\Pi}^{\mu\nu}.
\end{equation}
Here we have ``biased'' the matrix expression in Eq.~(\ref{eq:Coulomb_biasing})
into the simpler equation above. It assumes the form of an EM response
tensor in the presence of phase fluctuations, as in Eq.~(\ref{eq:phase_scre}),
but now the EM polarization tensors are substituted by their Coulomb-screened
versions: $Q^{\mu\nu}\to\widetilde{Q}^{\mu\nu}$. This expression
is manifestly gauge invariant as in Eq.~(\ref{eq:phase_scre}). Interestingly,
this biasing process can easily be done in the reverse manner. In
performing similar manipulations to arrive at Eq.~(\ref{eq:Coulomb_biasing}),
if we had first exchanged $Q^{\mu\nu}$ for $\Pi^{\mu\nu}$, instead
of $\widetilde{Q}^{\mu\nu}$, then it is a simple exercise to show
that by an analogue SVD the EM response tensor obtained reads
\begin{equation}
K_{\mathrm{mf}}^{\mu\nu}=\Pi^{\mu\nu}+\frac{\Pi_{0}^{\mu0}V\Pi^{0\nu}}{1-V\Pi^{00}}.
\end{equation}
This expression assumes a Coulomb-screened form, where each tensor
participating has been replaced by its ``phase-screened'' version:
$Q^{\mu\nu}\rightarrow\Pi^{\mu\nu}$. Evidently, since each $\Pi^{\mu\nu}$
is gauge invariant by itself, the whole expression above is gauge
invariant again. Naturally, both expressions for $K_{\mathrm{mf}}^{\mu\nu}$
above are equivalent.

Thus, we have introduced a process of folding the effects of each
fluctuating field via an SVD of the response tensors. This process
clearly biases the form of $K_{\mathrm{mf}}^{\mu\nu}$, although it
brings simplification. The denominators of the final form of these
response tensors contain the polaritonic resonances of the dielectric
functions~\citep{Kulik1981,Guo_Chien_He_2013,Garate_2013}. Equating
the two denominators equal to zero
\begin{equation}
q_{\lambda}\widetilde{Q}^{\lambda\sigma}q_{\sigma}=0=1-V\Pi^{00},
\end{equation}
one obtains the well-known Carlson-Goldman (CG) mode~\citep{Carlson_Goldman_1973,Carlson_Goldman_1975,Kulik1981},
where plasmons dress the phase fluctuation poles, gapping the phase
modes of charged superconductors. At $T=0$ this results in solely
a (double) plasmon mode, whereas in the vicinity of $T\sim T_{c}$
there is a soft mode (which was originally~\citep{Carlson_Goldman_1973,Carlson_Goldman_1975}
termed the CG mode) and a plasmon mode~\citep{Kulik1981}. Note that
the exact relation between the two denominators is: $q_{\lambda}\widetilde{Q}^{\lambda\sigma}q_{\sigma}\left(1-VQ^{00}\right)=q_{\lambda}Q^{\lambda\sigma}q_{\sigma}\left(1-V\Pi^{00}\right)$.

\subsection{EM response for superconductors (with amplitude fluctuations)\label{subsec:ChargedSCPhaseAmplitude}}

Returning to Eq.~(\ref{eq:SC_GF}), we now include the fluctuations
in $\rho\left(x\right)$. Contrary to the phase and Coulomb responses,
the amplitude part cannot be written solely in terms of the unscreened
EM response bubble $Q^{\mu\nu}\left(q\right)$. The additional objects
which must be defined for calculating the EM response functions read
as follows
\begin{align}
\left.\frac{\delta^{2}S_{\textrm{eff}}\left[{\bf \Delta},A\right]}{\delta\rho\left(-q\right)\delta\rho\left(q\right)}\right|_{A=0,\boldsymbol{\Delta}=\boldsymbol{\Delta}_{\text{mf}}\left[0\right]} & \equiv S^{\rho\rho}\left(q\right),\\
\left.\frac{\delta^{2}S_{\textrm{eff}}\left[{\bf \Delta},A\right]}{\delta\rho\left(-q\right)\delta\theta\left(q\right)}\right|_{A=0,\boldsymbol{\Delta}=\boldsymbol{\Delta}_{\text{mf}}\left[0\right]} & \equiv S^{\rho\theta}\left(q\right)=iq_{\beta}R^{\rho\beta}\left(q\right),
\end{align}
and similarly

\begin{align}
\left.\frac{\delta^{2}S_{\textrm{eff}}\left[{\bf \Delta},A\right]}{\delta\rho\left(-q\right)\delta A_{\mu}\left(q\right)}\right|_{A=0,\boldsymbol{\Delta}=\boldsymbol{\Delta}_{\text{mf}}\left[0\right]} & \equiv R^{\rho\mu}\left(q\right),\\
\left.\frac{\delta^{2}S_{\textrm{eff}}\left[{\bf \Delta},A\right]}{\delta\rho\left(-q\right)\delta\varphi\left(q\right)}\right|_{A=0,\boldsymbol{\Delta}=\boldsymbol{\Delta}_{\text{mf}}\left[0\right]} & \equiv S^{\rho\varphi}\left(q\right)=iR^{\rho0}\left(q\right).
\end{align}

Also note that just as $V^{-1}\left(q\right)$ contributed to $S^{\varphi\varphi}\left(q\right)$
{[}c.f. Eq.~(\ref{eq:Sfifi}){]}, the ``mass'' contribution for
$\rho\left(x\right)$ in the Hubbard-Stratonovich field in Eq.~(\ref{eq:SC_bos_act})
implies that $\widetilde{g}^{-1}$ contributes to $S^{\rho\rho}\left(q\right)$.

The EM response tensor now becomes

\begin{widetext}

\begin{equation}
K_{\text{mf}}^{\mu\nu}\left(q\right)=Q^{\mu\nu}-\left(\begin{array}{ccc}
R^{\mu\rho} & iQ^{\mu0} & iQ^{\mu\beta}\end{array}q_{\beta}\right)\left(\begin{array}{ccc}
S^{\rho\rho} & iR^{\rho0} & iR^{\rho\beta}q_{\beta}\\
iR^{0\rho} & V^{-1}-Q^{00} & -Q^{0\beta}q_{\beta}\\
-iq_{\alpha}R^{\alpha\rho} & q_{\alpha}Q^{\alpha0} & q_{\alpha}Q^{\alpha\beta}q_{\beta}
\end{array}\right)^{-1}\left(\begin{array}{c}
R^{\rho\nu}\\
iQ^{0\nu}\\
-iq_{\alpha}Q^{\alpha\nu}
\end{array}\right).
\end{equation}
\end{widetext}The SVD approach can also be implemented for this situation.
First note that, with some manipulation and SVD biasing, the determinant
can be reduced to two possible forms. The calculation is outlined
in Appendix~\ref{sec:Det_SVD} and results in 

\begin{align}
\det S\left(q\right) & =\left(V^{-1}-\overline{Q}^{00}\right)S^{\rho\rho}q_{\alpha}\widetilde{\overline{Q}}^{\alpha\beta}q_{\beta}\nonumber \\
 & =\left(V^{-1}-Q^{00}\right)\widetilde{S}^{\rho\rho}q_{\alpha}\overline{\widetilde{Q}}^{\alpha\beta}q_{\beta},\label{eq:detS}
\end{align}
where tilde variables are screened as in Eq.~(\ref{eq:Coulomb_scre});
for example, 

\begin{equation}
\widetilde{S}^{\rho\rho}=S^{\rho\rho}+\frac{R^{\rho0}VR^{0\rho}}{1-VQ^{00}}.
\end{equation}
Similarly, the process of ``folding'' the amplitude fluctuations
also leads to ``screened''-like tensors -- the ones with bars on
top. Repeating the calculation in Eq.~(\ref{eq:Coulomb_scre}), now
with only the amplitude contributions, one verifies,
\begin{equation}
\overline{Q}^{\alpha\beta}\equiv Q^{\alpha\beta}-\frac{R^{\alpha\rho}R^{\rho\beta}}{S^{\rho\rho}}.\label{eq:Qbar}
\end{equation}
Finally, tensors with both bars and tildes are interpreted to mean
first evaluate the tensors with respect to the outer screening symbol
and then with respect to the inner screening type. To be concrete,
as an example we have 
\begin{equation}
\overline{\widetilde{Q}}^{\alpha\beta}=\widetilde{Q}^{\alpha\beta}-\frac{\widetilde{R}^{\alpha\rho}\widetilde{R}^{\rho\beta}}{\widetilde{S}^{\rho\rho}}.
\end{equation}
From the two ways of writing the determinant above, and noticing that
\begin{align}
 & \left(V^{-1}-Q^{00}\right)S^{\rho\rho}+R^{\rho0}R^{0\rho}\nonumber \\
= & \left(V^{-1}-\overline{Q}^{00}\right)S^{\rho\rho}\nonumber \\
= & \left(V^{-1}-Q^{00}\right)\widetilde{S}^{\rho\rho},
\end{align}
we find an important identity:

\begin{equation}
q_{\alpha}\widetilde{\overline{Q}}^{\alpha\beta}q_{\beta}=q_{\alpha}\overline{\widetilde{Q}}^{\alpha\beta}q_{\beta}.\label{eq:QtildeQtilde}
\end{equation}
Using these expressions we can now perform the SVD process as in the
previous sections, the only requirement is to choose a biasing order
in which we want to take into account the influence of each type of
fluctuation. For example, taking into account the inversion of the
matrix $S\left(q\right)$ and the determinant above, we obtain

\begin{widetext}

\begin{align}
K_{\text{mf}}^{\mu\nu}\left(q\right)= & Q^{\mu\nu}-\frac{q_{\alpha}q_{\beta}}{q_{\lambda}\widetilde{\overline{Q}}^{\lambda\sigma}q_{\sigma}}\nonumber \\
 & \times\left(\begin{array}{c}
R^{\mu\rho}\\
Q^{\mu0}\\
Q^{\mu\beta}
\end{array}\right)^{T}\left(\begin{array}{ccc}
\frac{\left(V^{-1}-Q^{00}\right)\widetilde{Q}^{\alpha\beta}}{\left(V^{-1}-\overline{Q}^{00}\right)S^{\rho\rho}} & \frac{Q^{\alpha\beta}R^{\rho0}-Q^{\alpha0}R^{\rho\beta}}{\left(V^{-1}-\overline{Q}^{00}\right)S^{\rho\rho}} & -\frac{\left(V^{-1}-Q^{00}\right)\widetilde{R}^{\rho\beta}}{\left(V^{-1}-\overline{Q}^{00}\right)S^{\rho\rho}}\\
\frac{Q^{\alpha\beta}R^{0\rho}-R^{\alpha\rho}Q^{0\beta}}{\left(V^{-1}-\overline{Q}^{00}\right)S^{\rho\rho}} & -\frac{\overline{Q}^{\alpha\beta}}{\left(V^{-1}-\overline{Q}^{00}\right)} & \frac{\overline{Q}^{0\beta}}{\left(V^{-1}-\overline{Q}^{00}\right)}\\
-\frac{\left(V^{-1}-Q^{00}\right)\widetilde{R}^{\alpha\rho}}{\left(V^{-1}-\overline{Q}^{00}\right)S^{\rho\rho}} & \frac{\overline{Q}^{\alpha0}}{\left(V^{-1}-\overline{Q}^{00}\right)} & 1
\end{array}\right)\left(\begin{array}{c}
R^{\rho\nu}\\
Q^{0\nu}\\
Q^{\alpha\nu}
\end{array}\right).
\end{align}
Now we focus on the first term $Q^{\mu\nu}$. Introducing the effects
of amplitude fluctuations first (``bar'' variables) and subsequently
the regular screening from Coulomb fluctuations (``tilde'' variables),
a straightforward calculation and simplification using the relations
in Eq.~(\ref{eq:QtildeQtilde}) results in

\begin{align}
K_{\text{mf}}^{\mu\nu}= & \widetilde{\overline{Q}}^{\mu\nu}-\frac{q_{\alpha}q_{\beta}}{q_{\lambda}\widetilde{\overline{Q}}^{\lambda\sigma}q_{\sigma}}\nonumber \\
 & \times\left(\begin{array}{c}
R^{\mu\rho}\\
Q^{\mu0}\\
Q^{\mu\beta}
\end{array}\right)^{T}\left(\begin{array}{ccc}
\frac{\widetilde{R}^{\alpha\rho}\widetilde{R}^{\rho\beta}}{\left(\widetilde{S}^{\rho\rho}\right)^{2}} & -\frac{\overline{Q}^{\alpha0}\widetilde{R}^{\rho\beta}}{\left(V^{-1}-\overline{Q}^{00}\right)\widetilde{S}^{\rho\rho}} & -\frac{\widetilde{R}^{\rho\beta}}{\widetilde{S}^{\rho\rho}}\\
-\frac{\widetilde{R}^{\alpha\rho}\overline{Q}^{0\beta}}{\left(V^{-1}-\overline{Q}^{00}\right)\widetilde{S}^{\rho\rho}} & \frac{\overline{Q}^{\alpha0}\overline{Q}^{0\beta}}{\left(V^{-1}-\overline{Q}^{00}\right)^{2}} & \frac{\overline{Q}^{0\beta}}{\left(V^{-1}-\overline{Q}^{00}\right)}\\
-\frac{\widetilde{R}^{\alpha\rho}}{\widetilde{S}^{\rho\rho}} & \frac{\overline{Q}^{\alpha0}}{\left(V^{-1}-\overline{Q}^{00}\right)} & 1
\end{array}\right)\left(\begin{array}{c}
R^{\rho\nu}\\
Q^{0\nu}\\
Q^{\alpha\nu}
\end{array}\right).
\end{align}

\end{widetext}

This matrix is now of the form
\begin{equation}
M=\left(\begin{array}{ccc}
ab & -ad & -a\\
-bc & cd & c\\
-b & d & 1
\end{array}\right),
\end{equation}
where
\begin{eqnarray}
a=\frac{\widetilde{R}^{\rho\beta}}{\widetilde{S}^{\rho\rho}} & ,\  & b=\frac{\widetilde{R}^{\alpha\rho}}{\widetilde{S}^{\rho\rho}},\nonumber \\
c=\frac{\overline{Q}^{0\beta}}{V^{-1}-\overline{Q}^{00}} & ,\  & d=\frac{\overline{Q}^{\alpha0}}{V^{-1}-\overline{Q}^{00}}.
\end{eqnarray}

It displays two linearly dependent rows, thus suggesting the singular-value
decomposition. Performing the SVD and simplifying the result gives

\begin{align}
K_{\text{mf}}^{\mu\nu} & =\widetilde{\overline{Q}}^{\mu\nu}-\frac{\widetilde{\overline{Q}}^{\mu\beta}q_{\beta}q_{\alpha}\widetilde{\overline{Q}}^{\alpha\nu}}{q_{\lambda}\widetilde{\overline{Q}}^{\lambda\sigma}q_{\sigma}}\equiv\widetilde{\overline{\Pi}}^{\mu\nu}.\label{eq:all_in}
\end{align}
Setting $q_{\lambda}\widetilde{\overline{Q}}^{\lambda\sigma}q_{\sigma}=0$
gives the collective mode dispersion for the polaritons induced by
simultaneous Coulomb, phase, and amplitude fluctuations. Again, gauge
invariance in the SVD-simplified EM response in Eq.~(\ref{eq:all_in})
is manifest:
\begin{align}
q_{\mu}K_{\text{mf}}^{\mu\nu} & =q_{\mu}\widetilde{\overline{Q}}^{\mu\nu}-\frac{q_{\mu}\widetilde{\overline{Q}}^{\mu\beta}q_{\beta}q_{\alpha}\widetilde{\overline{Q}}^{\alpha\nu}}{q_{\lambda}\widetilde{\overline{Q}}^{\lambda\sigma}q_{\sigma}}=0.
\end{align}
As in the previous section, other equivalent forms for the EM response
can be obtained by reversing the order in the SVD processes. For example,
$K_{\text{mf}}^{\mu\nu}=\widetilde{\overline{\Pi}}^{\mu\nu}=\overline{\widetilde{\Pi}}^{\mu\nu}$.

\section{The Meissner effect in the presence of collective modes \label{sec:Meissner}}

\subsection{Kubo formula}

In this section we calculate the superfluid density for superfluid
and superconducting systems with amplitude, phase, and Coulomb fluctuations
incorporated. It was shown in the previous section that the EM response
for a system with all these three types of fluctuations can be compactly
written as in Eq.~(\ref{eq:all_in}). Here we will use this formula
to study the Meissner response for both $s$- and $p$-wave systems.
The Kubo formula for the superfluid density tensor is~\citep{AGDBook}
\begin{equation}
\frac{e^{2}}{m}n_{s}^{xx}=\lim_{{\bf q}\rightarrow0}K_{\text{mf}}^{ii}\left(\Omega=0,{\bf q}\right),
\end{equation}
with no implicit index summation. It is crucial that the static limit,
$\Omega=0$, is taken before the long-wavelength limit ${\bf q}\rightarrow0$
is considered. This particular order of limits is appropriate for
a thermodynamic quantity, whereas the converse procedure is apt for
the calculation of optical properties, namely the DC electrical conductivity
for instance. For nonuniform systems, the limit ${\bf q}\rightarrow0$
must also be carefully specified. To ascertain the appropriate definition,
recall that in the presence of an external EM vector potential $A_{\nu}$,
the EM current is $J^{\mu}\left(x\right)=\int_{x^{\prime}}K^{\mu\nu}\left(x,x^{\prime}\right)A_{\nu}\left(x^{\prime}\right)$.
The continuity equation is $\partial_{\mu}J^{\mu}=0$; this statement
enforces conservation of global particle number (global U(1) symmetry)
for a neutral superfluid, whereas for a charged system it enforces
conservation of charge. In terms of the response kernel, this equation
becomes $\left(\partial_{\mu}K^{\mu\nu}\right)A_{\nu}=0.$ The solution
to this equation, for an arbitrary $A_{\nu}$, is to require a gauge-invariant
EM response: $\partial_{\mu}K^{\mu\nu}=0$, which in momentum space
reads $q_{\mu}K^{\mu\nu}=0$. As shown in the previous section, the
SVD approach enables this to be manifestly satisfied.

To compute $n_{s}$ it is convenient to work in the gauge where $\partial_{\mu}A_{\mu}=0$
(Lorenz gauge), which reduces to the Coulomb gauge $\nabla\cdot{\bf A}=0$
in the static limit. The momentum-space form of the Coulomb gauge
is ${\bf q}\cdot{\bf A}=0$. In deriving the superfluid density $n_{s}^{ii}$,
only the $i$th component of the vector field must be non-vanishing:
$A_{i}\neq0$. The Coulomb gauge condition then reduces to $q_{i}A_{i}=0$,
demanding $q_{i}=0$. The other momentum components go to zero only
in the limit. Thus, the appropriate Kubo formula for $n_{s}^{ii}$
is
\begin{equation}
\frac{e^{2}}{m}n_{s}^{xx}=\lim_{q_{k}\neq q_{i}\rightarrow0,q_{i}=0}K_{\text{mf}}^{ii}\left(\Omega=0,{\bf q}\right).
\end{equation}
This Kubo formula explains why the superfluid density is often termed
a ``transverse'' response~\citep{AGDBook,PinesNozieres1}. In the
particular case of nonuniform superfluids, however, the appellation
``transverse'' loses its significance. The importance of computing
the superfluid density in the appropriate limiting fashion was discussed
in\textcolor{red}{{} }Ref.~\citep{Boyack_2017}, where it was shown
that for the Fulde-Ferrell superfluid the amplitude collective mode
contributes to the superfluid density. A general argument for why
collective modes do not need to be considered in the superfluid density
response of uniform superfluids is as follows~\citep{AGDBook}. In
the presence of the external vector potential $A$, the order parameter
can be expanded to quadratic order in $A$ as

\begin{equation}
\Delta\left[A\right]=\Delta\left[A=0\right]+\Delta^{\left(1\right)}\left[A\right]+\mathcal{O}\left(A^{2}\right),
\end{equation}
Since the order parameter $\Delta$ is a scalar, whereas the vector
potential $A$ is a vector, $\Delta$ can depend only on scalar-valued
functions of $A$. For a uniform superfluid, the only such scalar
quantity is $\nabla\cdot{\bf A}$. In the Coulomb gauge, where $\nabla\cdot{\bf A}=0$,
it follows that $\Delta^{\left(1\right)}=0$. Thus, collective modes
do not contribute to the superfluid density in a uniform superfluid.
In the case of a nonuniform superfluid, there are potentially other
scalar quantities that depend on ${\bf A}$ and thus $\Delta^{\left(1\right)}$
need not be zero. The next section provides an explicit calculation
of the superfluid density for $s$- and $p$-wave superfluids with
amplitude, phase, and Coulomb interactions. 

\subsection{Explicit superfluid density calculation}

First consider the case of a uniform $s$-wave superfluid. Without
loss of generality, since the system is uniform we only need to study
the response in one direction, say $\hat{x}$. Using the formalism
developed in the previous sections, the superfluid density is given
by 
\begin{align}
\frac{e^{2}}{m}n_{s}^{xx} & =\lim_{q_{x}=0,q_{y}\rightarrow0}\left[\widetilde{\overline{Q}}^{xx}-\frac{\widetilde{\overline{Q}}^{xi}q_{i}q_{j}\widetilde{\overline{Q}}^{jx}}{q_{k}\widetilde{\overline{Q}}^{kl}q_{l}}\right]\nonumber \\
 & =\lim_{q_{x}=0,q_{y}\rightarrow0}\left[\widetilde{\overline{Q}}^{xx}-\frac{\widetilde{\overline{Q}}^{xy}\widetilde{\overline{Q}}^{yx}}{\widetilde{\overline{Q}}^{yy}}\right].\label{eq:Rhosxx}
\end{align}
In the small-momentum limit, $R^{\rho j}\left(0,\mathbf{q}\to0\right)=0$;
this is because in this limit the tensor structure requires $R^{\rho j}\left(0,{\bf q}\rightarrow0\right)\sim q^{j}\rightarrow0$.
Thus, the generalized response functions are 
\begin{align}
\widetilde{\overline{Q}}^{xj} & =\overline{Q}^{xj}+\frac{\overline{Q}^{x0}\overline{Q}^{0j}}{V^{-1}-\overline{Q}^{00}}=Q^{xj}.
\end{align}
As a result, the superfluid density is 
\begin{equation}
\frac{e^{2}}{m}n_{s}^{xx}=\lim_{q_{x}=0,q_{y}\rightarrow0}Q^{xx}.
\end{equation}
This proves that without any particular assumptions about particle-hole
symmetry, i.e., whether or not the amplitude and Coulomb mode decouple
$\left(R^{\rho0}\neq0\right)$~\citep{Levin_2000}, the superfluid
density for an $s$-wave system has no contributions from amplitude,
phase, or Coulomb collective modes. This is an explicit proof of the
argument presented in the previous section. 

Now consider a spinless-$\left(p+ip\right)$ superfluid in two spatial
dimensions. The $x$ and $y$ responses are equivalent, thus we again
need only to consider the former. The superfluid density is as given
in Eq.~(\ref{eq:Rhosxx}). Again $R^{\rho j}\left(0,\mathbf{q}\to0\right)=0$
remains true, and thus 
\begin{equation}
\frac{e^{2}}{m}n_{s}^{xx}=\lim_{q_{x}=0,q_{y}\rightarrow0}Q^{xx}.
\end{equation}
This particular limit is computed as shown below. After performing
the Matsubara frequency summation, the response function is~\citep{Yakovenko_2008,Guo_Chien_He_2013}: 

\begin{widetext}
\begin{align}
Q^{ij}\left(i\Omega_{m},{\bf q}\right) & =\frac{e^{2}}{2}\int\frac{d^{2}\mathbf{p}}{\left(2\pi\right)^{2}}\frac{\mathbf{p}^{i}}{m}\frac{\mathbf{p}^{j}}{m}\left[\left(1+\frac{\xi_{\mathbf{p}}^{+}\xi_{\mathbf{p}}^{-}+\Delta_{0}^{2}\mathbf{p}_{+}\cdot\mathbf{p}_{-}/p_{F}^{2}}{E_{\mathbf{p}}^{+}E_{\mathbf{p}}^{-}}\right)\frac{E_{\mathbf{p}}^{+}-E_{\mathbf{p}}^{-}}{\left(E_{\mathbf{p}}^{+}-E_{\mathbf{p}}^{-}\right)^{2}-\left(i\Omega_{m}\right)^{2}}\left[f\left(E_{\mathbf{p}}^{+}\right)-f\left(E_{\mathbf{p}}^{-}\right)\right]\right.\nonumber \\
 & \left.-\left(1-\frac{\xi_{\mathbf{p}}^{+}\xi_{\mathbf{p}}^{-}+\Delta_{0}^{2}\mathbf{p}_{+}\cdot\mathbf{p}_{-}/p_{F}^{2}}{E_{\mathbf{p}}^{+}E_{\mathbf{p}}^{-}}\right)\frac{E_{\mathbf{p}}^{+}+E_{\mathbf{p}}^{-}}{\left(E_{\mathbf{p}}^{+}+E_{\mathbf{p}}^{-}\right)^{2}-\left(i\Omega_{m}\right)^{2}}\left[1-f\left(E_{\mathbf{p}}^{-}\right)-f\left(E_{\mathbf{p}}^{+}\right)\right]\right]+\frac{ne^{2}}{m}\delta^{ij},
\end{align}
\end{widetext}where ${\bf p}_{\pm}={\bf p}\pm{\bf q}/2$, $\xi_{\mathbf{p}}^{\pm}\equiv\xi_{\mathbf{p}\pm\mathbf{q}/2},E_{\mathbf{p}}^{\pm}\equiv E_{\mathbf{p}\pm\mathbf{q}/2}$,
with $\xi_{\mathbf{p}}=\mathbf{p}^{2}/\left(2m\right)-\mu,E_{\mathbf{p}}=\sqrt{\xi_{\mathbf{p}}^{2}+\Delta_{0}^{2}\mathbf{p}^{2}/p_{F}^{2}}$,
and $n$ is the total number density. Taking the appropriate frequency
and momentum limits results in
\begin{align}
\frac{e^{2}}{m}n_{s}^{xx}= & e^{2}\left[\frac{n}{m}+\int\frac{d^{2}\mathbf{p}}{\left(2\pi\right)^{2}}\left(\frac{p^{x}}{m}\right)^{2}\frac{\partial f\left(E_{\mathbf{p}}\right)}{\partial E_{\mathbf{p}}}\right]\nonumber \\
= & e^{2}\int\frac{d^{2}\mathbf{p}}{\left(2\pi\right)^{2}}\left(\frac{p^{x}}{m}\right)^{2}\frac{\Delta_{0}^{2}\mathbf{p}^{2}/p_{F}^{2}}{E_{\mathbf{p}}^{2}}\nonumber \\
 & \times\left[\frac{1-2f\left(E_{\mathbf{p}}\right)}{2E_{\mathbf{p}}}+\frac{\partial f\left(E_{\mathbf{p}}\right)}{\partial E_{\mathbf{p}}}\right].
\end{align}
In general, for a superfluid system with only one external momentum,
namely the momentum ${\bf q}$ of the external vector potential ${\bf A}$,
the EM response can be decomposed into terms comprised of $\delta^{ij}$
and $q^{i}q^{j}/{\bf q}^{2}$. In the limit ${\bf q}\rightarrow0$,
as defined above, it \textcolor{black}{follows} that the off-diagonal
terms vanish and thus the superfluid density reduces to the standard
undressed bubble term. \textcolor{black}{Unless there are other $\textit{external}$
vectors that can couple to the vector potential, the superfluid density
always reduces to the undressed bubble term. This statement is a generalization
of the analysis in the previous section, which considered only uniform
superfluids; here we extend the veracity of the previous proof to
include all kinds of superfluids without other $\textit{external}$
vectors that couple to the vector potential. }

\subsection{Transverse and longitudinal responses}

In Ref.~\citep{Millis_1987} the EM response for nonuniform superfluids
without amplitude fluctuations was derived. This particular article
highlighted that for such superfluids the collective modes are, in
general, no longer solely ``longitudinal'', and moreover these modes
can be important in what are conventionally termed ``transverse''
response functions in the case of uniform systems. In this section
we show that our generalized formula reproduces the particular case
considered in Ref.~\citep{Millis_1987}, namely, a neutral system
with only phase fluctuations of the order parameter. Using Eq.~(\ref{eq:all_in}),
the response function for such a system, in the static limit, is given
by
\begin{equation}
K_{\text{mf}}^{ij}\left(0,{\bf q}\right)=Q^{ij}\left(0,{\bf q}\right)-\frac{Q^{ia}\left(0,{\bf q}\right)q_{a}q_{b}Q^{bj}\left(0,{\bf q}\right)}{q_{c}Q^{cd}\left(0,{\bf q}\right)q_{d}}.\label{eq:Kresp}
\end{equation}
The undressed EM response (for a spin-$\frac{1}{2}$ system with $e=1$)
reads~\citep{AGDBook,Guo_Chien_He_2013}
\begin{align}
Q^{ij}\left(0,{\bf q}\right)= & 2\sum_{p}\left(\frac{p^{i}}{m}\frac{p^{j}}{m}\right)\left[G\left(i\omega_{n},{\bf p}_{+}\right)G\left(i\omega_{n},{\bf p}_{-}\right)\right.\nonumber \\
 & \left.+F^{*}\left(i\omega_{n},{\bf p}_{+}\right)F\left(i\omega_{n},{\bf p}_{-}\right)\right]+\frac{n}{m}\delta^{ij},\label{eq:QResp}
\end{align}
The non-bold momenta are four-vectors $p^{\mu}=\left(i\omega_{n},{\bf p}\right)$
with $\omega_{n}$ a fermionic Matsubara frequency. For simplicity,
let us focus on a system with a general momentum-angle-dependent gap
$\Delta_{\mathbf{p}}\equiv\Delta\left(\hat{\mathbf{p}}\right)$. The
single-particle and anomalous Green's functions are~\citep{AGDBook,Guo_Chien_He_2013}
\begin{eqnarray}
G\left(i\omega_{n},{\bf p}\right) & = & -\frac{i\omega_{n}+\xi_{\mathbf{p}}}{\omega_{n}^{2}+\xi_{\mathbf{p}}^{2}+\left|\Delta_{\mathbf{p}}\right|^{2}},\\
F\left(i\omega_{n},{\bf p}\right) & = & \frac{\Delta_{\mathbf{p}}}{\omega_{n}^{2}+\xi_{\mathbf{p}}^{2}+\left|\Delta_{\mathbf{p}}\right|^{2}}.
\end{eqnarray}
A generic static correlation function for a uniform system has the
form 
\begin{equation}
K^{ij}\left(0,{\bf q}\right)=\chi_{L}\frac{q^{i}q^{j}}{\mathbf{q}^{2}}+\chi_{T}\left(\delta^{ij}-\frac{q^{i}q^{j}}{\mathbf{q}^{2}}\right).\label{eq:Corr_Func}
\end{equation}
Here, $\chi_{T}$ and $\chi_{L}$ denote the transverse and longitudinal
part of the full response function, respectively. By taking the dot
product with $q^{i}$ and $q^{j}$, the longitudinal part is 
\begin{equation}
\chi_{L}=\frac{q^{i}K^{ij}q^{j}}{q^{2}}.
\end{equation}
The longitudinal part of the total response gives zero contribution
to the Meissner effect: the full response is purely transverse. In
the small-momentum limit the collective-mode part of the response
(the second term in Eq.~(\ref{eq:Kresp})) is purely longitudinal,
and thus it gives zero contribution to the superfluid density. 

Let $i=j$ in Eq.~(\ref{eq:Corr_Func}) and take the trace to obtain
$\sum_{i}K^{ii}=\chi_{L}+2\chi_{T}.$ Therefore the transverse part
is 
\begin{equation}
\chi_{T}=\frac{1}{2}\left(\sum_{i}K^{ii}-\chi_{L}\right).
\end{equation}
Let $\left(m/n\right)\chi_{T}\equiv\chi_{T}^{\prime}$. Using Eq.~(\ref{eq:QResp}),
this becomes 
\begin{align}
\chi_{T}^{\prime}\left(q\right)= & \frac{1}{mn}\sum_{p}p^{2}\sin^{2}\left(\theta\right)\left[G\left(i\omega_{n},{\bf p}_{+}\right)G\left(i\omega_{n},{\bf p}_{-}\right)\right.\nonumber \\
 & \left.+F^{*}\left(i\omega_{n},{\bf p}_{+}\right)F\left(i\omega_{n},{\bf p}_{-}\right)\right]+1.
\end{align}
We drop the $q$ dependence in the argument of $\chi_{T}$ from now
on. To evaluate this quantity we invoke standard Fermi-liquid theory
and assume a constant density of states near the Fermi-surface. Using
this approximation, the transverse response then becomes~\citep{AGDBook}
\begin{align}
\chi_{T}^{\prime}= & 1+T\frac{3}{4}\sum_{\omega_{n}}\int_{0}^{\pi}d\theta\sin^{3}(\theta)\int_{-\infty}^{\infty}d\xi\nonumber \\
 & \times\frac{\left(i\omega_{n}+\xi_{+}\right)\left(i\omega_{n}+\xi_{-}\right)+\left|\Delta_{\mathbf{p}}\right|^{2}}{\left(\omega_{n}^{2}+\xi_{+}^{2}+\left|\Delta_{\mathbf{p}}\right|^{2}\right)\left(\omega_{n}^{2}+\xi_{-}^{2}+\left|\Delta_{\mathbf{p}}\right|^{2}\right)}.
\end{align}
Here, $\xi_{\pm}=\xi\pm\frac{1}{2}qv_{F}\cos\left(\theta\right)$
with $v_{F}=p_{F}/m$ the Fermi speed, and we have also used $k_{F}^{3}=3\pi^{2}n$.
As discussed in Ref.~\citep{AGDBook}, the result of performing the
Matsubara frequency summation followed by the $\xi$ integration leads
to the correct normal-state result. However, performing this procedure
in the reverse order leads to a different answer, in contradiction
to the absence of a normal-state Meissner effect. To circumvent this
problem, the method employed is to add and subtract the normal-state
density expression. This enables performing the integration over $\xi$
first, which results in 
\begin{equation}
\chi_{T}^{\prime}=\frac{3\pi}{4}T\sum_{\omega_{n}}\int_{-1}^{1}\frac{dx}{\sqrt{\omega_{n}^{2}+\left|\Delta_{\mathbf{p}}\right|^{2}}}\frac{\left(1-x^{2}\right)\left|\Delta_{\mathbf{p}}\right|^{2}}{\omega_{n}^{2}+\left|\Delta_{\mathbf{p}}\right|^{2}+\frac{1}{4}q^{2}v_{F}^{2}x^{2}}.\label{eq:Transverse_Resp}
\end{equation}
For comparison, the EM current given in Ref.~\citep{Millis_1987}
reads

\begin{equation}
{\bf J}\left({\bf q}\right)=\int dS_{p}R\left(\hat{{\bf p}}\right)\hat{{\bf p}}\left[\hat{{\bf p}}\cdot{\bf A}\left({\bf q}\right)-\hat{{\bf p}}\cdot\hat{{\bf q}}\phi\left(q\right)\right],
\end{equation}
with the function $R\left(\hat{{\bf p}}\right)\equiv R\left(\hat{{\bf p}};0,\hat{{\bf q}}\right)$
given by

\begin{equation}
R\left(\hat{{\bf p}};0,\hat{{\bf q}}\right)=T\sum_{\omega_{n}}\frac{1}{\sqrt{\omega_{n}^{2}+\left|\Delta_{\mathbf{p}}\right|^{2}}}\frac{\left|\Delta_{\mathbf{p}}\right|^{2}}{\omega_{n}^{2}+\left|\Delta_{\mathbf{p}}\right|^{2}+\frac{1}{4}q^{2}v_{F}^{2}x^{2}}
\end{equation}
and $\phi\left(q\right)$ given by
\begin{equation}
\phi\left(q\right)=\frac{\int dS_{l}R\left(\hat{{\bf l}}\right)\hat{{\bf l}}\cdot\hat{{\bf q}}\hat{{\bf l}}\cdot{\bf A}\left({\bf q}\right)}{\int dS_{k}R\left(\hat{{\bf k}}\right)\left(\hat{{\bf k}}\cdot\hat{{\bf q}}\right)^{2}}.
\end{equation}
It is straightforward to check that this expression conserves particle
number: ${\bf q}\cdot{\bf J}=0$. The corresponding response kernel
is thus
\begin{eqnarray}
K^{ij}\left(\Omega=0,{\bf q}\right) & = & Q^{ij}-\frac{Q^{ia}q_{a}q_{b}Q^{bj}}{q_{c}Q^{cd}q_{d}},
\end{eqnarray}
where $Q^{ij}\left(0,{\bf q}\right)\equiv\int dS_{p}\hat{p}^{i}R\left(\hat{{\bf p}};0,{\bf q}\right)\hat{p}^{j}$,
with $dS_{p}$ the measure on the Fermi surface. Furthermore, the
transverse part of the response is~\footnote{In Ref.~\citep{Millis_1987} the prefactor of $1/m$ in the EM vertices
was omitted. Furthermore, the explicit form for $dS_{p}$ was unspecified.
The definition given in Ref.~\citep{Millis_1987} is that, up to
a constant, it is the angle-dependent density of states. Using this
definition, we accordingly find $\int dS_{p}\sin^{2}\left(\theta\right)f\left(\theta\right)\sim\frac{mp_{F}}{\pi^{2}}p_{F}^{2}\int\frac{d\theta d\phi}{4\pi}\sin^{3}\left(\theta\right)f\left(\theta\right)=\frac{3}{2}\frac{n}{m}m^{2}\int dx\left(1-x^{2}\right)f\left(x\right)$.
The factor of $m^{2}$ drops out once the vertices are appropriately
restored. There is an additional factor of $\pi$ that also must be
restored. Nevertheless, our result is in exact agreement with Eq.~(37.15)
of Ref.~\citep{AGDBook} for the $s$-wave case (accounting for the
differences in definition of response kernel.). } 
\begin{align}
\chi_{T}^{\prime} & =\frac{3\pi}{4}\int_{-1}^{1}dx\left(1-x^{2}\right)R\left(\hat{{\bf p}};0,{\bf q}\right)\nonumber \\
 & =\frac{3\pi}{4}T\sum_{\omega_{n}}\int_{-1}^{1}\frac{dx}{\sqrt{\omega_{n}^{2}+\left|\Delta_{\mathbf{p}}\right|^{2}}}\frac{\left(1-x^{2}\right)\left|\Delta_{\mathbf{p}}\right|^{2}}{\omega_{n}^{2}+\left|\Delta_{\mathbf{p}}\right|^{2}+\frac{1}{4}q^{2}v^{2}x^{2}}.\label{eq:Transverse_Resp2}
\end{align}
Therefore, we have shown that Eq.~(\ref{eq:Transverse_Resp}), which
followed from our generalized formula for phase fluctuations, agrees
with Eq.~(\ref{eq:Transverse_Resp2}). 

To finish, consider a two-dimensional superfluid where the current
and vector potential are parallel: $J^{x}=K^{xx}A_{x},J^{y}=K^{yy}A_{y}.$
The ratio of the EM kernels is 
\begin{equation}
\frac{\lambda_{x}^{2}}{\lambda_{y}^{2}}=\frac{K^{xx}}{K^{yy}}.
\end{equation}
In Ref.~\citep{Millis_1987}, where the effects from phase collective
modes were the focus, it was pointed out that in the case of a dipolar
superfluid this quantity is not unity. The analysis in this section
shows that, in the static and long-wavelength limit, the full response
is purely transverse, and thus there is no collective-mode contribution
to the above ratio. The reason for its departure from unity~\citep{Millis_1987}
is merely because the undressed bubble contributions are distinct
for the dipolar superfluid.

\section{Conclusions~\label{sec:Conclusions}}

The rich physics associated with superfluids and superconductors is
most perceptible in the collective fluctuations of the order parameter.
These modes show that superconductors are more than just gapped fluids
of condensed electron-electron pairs. Rather, superconductors are
systems replete with collective excitations due to coherent many-particle
effects. Historically these modes were first studied in the context
of restoring gauge invariance in a superconductor. More recently,
however, a bevy of literature has studied these excitations in more
general settings, and one particularly important problem has been
understanding their role in the Meissner effect. 

The antecedent literature to the present work suggested that collective 
modes may be ignored in $s$-wave systems, but must be accounted for if 
the order parameter is anisotropic ($p$-wave, $d$-wave, etc). In this paper 
we have extended this analysis by developing a general method for computing
the electromagnetic response in systems with multiple collective modes.
We have shown that, in fact, collective modes do not contribute to
the Meissner effect in neither uniform nor nonuniform superconductors.
An exception to this scenario comes about when external wavevector
scales exist, as in Fulde-Ferrell finite-momentum paired superconductors.
The by-product of our study was to show that through singular-value
decompositions, the electromagnetic response in a system with multiple
collective modes present can naturally be computed by folding the
various response tensors into dressed constituents. With all details
we provided, we anticipate that this methodology will also prove useful
in other contexts such as charge-density waves and quantum magnetism. 

\section{acknowledgments}

RB and PLSL contributed equally to this work. We thank Shinsei Ryu
and Joseph Maciejko for helpful discussions and suggestions. RB is
supported by the Theoretical Physics Institute at the University of
Alberta. PLSL is supported by the Canada First Research Excellence
Fund.

\begin{widetext}

\appendix

\section{Derivation of the mean-field EM response tensor~\label{sec:tensor}}

In this excursus we derive in detail the EM response tensor in Eq.~(\ref{eq:EM_resp}).
For concreteness, whenever we write $\delta S_{\text{eff}}\left[\boldsymbol{{\bf \Delta}},A\right]/\delta A_{\nu}\left(x^{\prime}\right)$
(no $\mathbf{A}$ dependence in $\boldsymbol{\Delta}$), we mean the
explicit $A$ dependence is being differentiated, with the collective-mode
fields fixed. The functional chain rule produces
\begin{eqnarray}
\frac{\delta S_{\text{eff}}\left[\boldsymbol{{\bf \Delta}}_{\textrm{mf}}\left[A\right],A\right]}{\delta A_{\nu}\left(y\right)} & = & \left(\frac{\delta S_{\text{eff}}\left[\boldsymbol{{\bf \Delta}},A\right]}{\delta A_{\nu}\left(y\right)}\right)_{\boldsymbol{{\bf \Delta}}_{\textrm{mf}}\left[A\right]}+\int_{z}\left(\frac{\delta S_{\text{eff}}\left[\boldsymbol{{\bf \Delta}},A\right]}{\delta{\bf \Delta}_{a}\left(z\right)}\right)_{\boldsymbol{{\bf \Delta}}_{\textrm{mf}}\left[A\right]}\frac{\delta{\bf \Delta}_{a}^{\textrm{mf}}\left[A\right]\left(z\right)}{\delta A_{\nu}\left(y\right)}.
\end{eqnarray}
At the end of the calculation the value of ${\bf \Delta}$ is set
to its mean-field value ${\bf \Delta}_{\textrm{mf}}\left[A\right].$
Similarly, the second derivative of the above expression reads
\begin{eqnarray}
\frac{\delta^{2}S_{\text{eff}}\left[\boldsymbol{{\bf \Delta}}_{\textrm{mf}}\left[A\right],A\right]}{\delta A_{\mu}\left(x\right)\delta A_{\nu}\left(y\right)} & = & \left(\frac{\delta^{2}S_{\text{eff}}\left[\boldsymbol{{\bf \Delta}},A\right]}{\delta A_{\mu}\left(x\right)\delta A_{\nu}\left(y\right)}\right)_{\boldsymbol{{\bf \Delta}}=\boldsymbol{{\bf \Delta}}_{\textrm{mf}}\left[A\right]}+\int_{z,z'}\frac{\delta{\bf \Delta}_{a}^{\textrm{mf}}\left[A\right]\left(z\right)}{\delta A_{\mu}\left(x\right)}\left(\frac{\delta^{2}S_{\text{eff}}\left[\boldsymbol{{\bf \Delta}},A\right]}{\delta{\bf \Delta}_{a}\left(z\right)\delta{\bf \Delta}_{b}\left(z^{\prime}\right)}\right)_{\boldsymbol{{\bf \Delta}}=\boldsymbol{{\bf \Delta}}_{\textrm{mf}}\left[A\right]}\frac{\delta{\bf \Delta}_{b}^{\textrm{mf}}\left[A\right]\left(z^{\prime}\right)}{\delta A_{\nu}\left(y\right)}\nonumber \\
 & + & \int_{z}\frac{\delta{\bf \Delta}_{a}^{\textrm{mf}}\left[A\right]\left(z\right)}{\delta A_{\mu}\left(x\right)}\left(\frac{\delta^{2}S_{\text{eff}}\left[\boldsymbol{{\bf \Delta}},A\right]}{\delta{\bf \Delta}_{a}\left(z\right)\delta A_{\nu}\left(y\right)}\right)_{\boldsymbol{{\bf \Delta}}=\boldsymbol{{\bf \Delta}}_{\textrm{mf}}\left[A\right]}+\int_{z}\left(\frac{\delta^{2}S_{\text{eff}}\left[\boldsymbol{{\bf \Delta}},A\right]}{\delta A_{\mu}\left(x\right)\delta{\bf \Delta}_{a}\left(z\right)}\right)_{\boldsymbol{{\bf \Delta}}=\boldsymbol{{\bf \Delta}}_{\textrm{mf}}\left[A\right]}\frac{\delta{\bf \Delta}_{a}^{\textrm{mf}}\left[A\right]\left(z\right)}{\delta A_{\nu}\left(y\right)}\nonumber \\
 & + & \int_{z}\left(\frac{\delta S_{\text{eff}}\left[\boldsymbol{{\bf \Delta}},A\right]}{\delta{\bf \Delta}_{a}\left(z\right)}\right)_{\boldsymbol{{\bf \Delta}}=\boldsymbol{{\bf \Delta}}_{\textrm{mf}}\left[A\right]}\frac{\delta^{2}{\bf \Delta}_{a}^{\textrm{mf}}\left[A\right]\left(z\right)}{\delta A_{\mu}\left(x\right)\delta A_{\nu}\left(y\right)}.\label{eq:EM_Response2}
\end{eqnarray}
Since we are interested in the mean-field EM response, we can invoke
the saddle-point condition 
\begin{eqnarray}
0 & = & \left.\frac{\delta S_{\text{eff}}\left[\boldsymbol{{\bf \Delta}},A\right]}{\delta{\bf \Delta}_{a}\left(z\right)}\right|_{\boldsymbol{{\bf \Delta}}=\boldsymbol{{\bf \Delta}}_{\textrm{mf}}\left[A\right]};\label{eq:SP_cond}
\end{eqnarray}
thus the last term in Eq.~(\ref{eq:EM_Response2}) gives zero mean-field
contribution and can be dropped. If one were to consider the EM response
at the Gaussian order, however, then this term would contribute. It
remains to compute the derivatives of the collective-mode fields ${\bf \Delta}_{a}$
with respect to the vector potential. This can be done by considering
the saddle-point conditions. Differentiating Eq.~(\ref{eq:SP_cond})
with respect to $A$ gives
\begin{eqnarray}
0 & = & \frac{\delta}{\delta A_{\nu}\left(y\right)}\left(\frac{\delta S_{\text{eff}}\left[\boldsymbol{{\bf \Delta}},A\right]}{\delta{\bf \Delta}_{a}\left(z\right)}\right)_{\boldsymbol{{\bf \Delta}}=\boldsymbol{{\bf \Delta}}_{\textrm{mf}}\left[A\right]}\nonumber \\
 & = & \left(\frac{\delta^{2}S_{\text{eff}}\left[\boldsymbol{{\bf \Delta}},A\right]}{\delta A_{\nu}\left(y\right)\delta{\bf \Delta}_{a}\left(z\right)}\right)_{\boldsymbol{{\bf \Delta}}=\boldsymbol{{\bf \Delta}}_{\textrm{mf}}\left[A\right]}+\int_{z'}\left(\frac{\delta^{2}S_{\text{eff}}\left[\boldsymbol{{\bf \Delta}},A\right]}{\delta{\bf \Delta}_{a}\left(z\right)\delta{\bf \Delta}_{b}\left(z^{\prime}\right)}\right)_{\boldsymbol{{\bf \Delta}}=\boldsymbol{{\bf \Delta}}_{\textrm{mf}}[A]}\frac{\delta{\bf \Delta}_{b}^{\textrm{mf}}\left[A\right]\left(z^{\prime}\right)}{\delta A_{\nu}\left(y\right)}.
\end{eqnarray}
Inverting the saddle-point integral equation yields
\begin{eqnarray}
\frac{\delta{\bf \Delta}_{b}^{\textrm{mf}}\left[A\right]\left(z^{\prime}\right)}{\delta A_{\nu}\left(y\right)} & = & -\int_{z}\left(\frac{\delta^{2}S_{\text{eff}}\left[\boldsymbol{{\bf \Delta}},A\right]}{\delta{\bf \Delta}_{b}\left(z^{\prime}\right)\delta{\bf \Delta}_{a}\left(z\right)}\right)_{\boldsymbol{{\bf \Delta}}=\boldsymbol{{\bf \Delta}}_{\textrm{mf}}\left[A\right]}^{-1}\left(\frac{\delta^{2}S_{\text{eff}}\left[\boldsymbol{{\bf \Delta}},A\right]}{\delta{\bf \Delta}_{a}\left(z\right)\delta A_{\nu}\left(y\right)}\right)_{\boldsymbol{{\bf \Delta}}=\boldsymbol{{\bf \Delta}}_{\textrm{mf}}\left[A\right]}.
\end{eqnarray}
Substituting this into Eq.~(\ref{eq:EM_Response2}) and taking $A\to0$,
we then obtain Eq.~(\ref{eq:EM_resp}) of the main text:

\begin{align}
K_{\textrm{mf}}^{\mu\nu}\left(x,y\right)= & \left(\frac{\delta^{2}S_{\text{eff}}\left[{\bf \Delta},A\right]}{\delta A_{\mu}\left(x\right)\delta A_{\nu}\left(y\right)}\right)_{{\bf \Delta}_{\textrm{mf}}[0]}\nonumber \\
 & -\int_{z,z'}\left(\frac{\delta^{2}S_{\text{eff}}\left[{\bf \Delta},A\right]}{\delta A_{\mu}\left(x\right)\delta{\bf \Delta}_{a}\left(z\right)}\right)_{{\bf \Delta}_{\textrm{mf}}[0]}\left(\frac{\delta^{2}S_{\text{eff}}\left[{\bf \Delta},A\right]}{\delta{\bf \Delta}_{a}\left(z\right)\delta{\bf \Delta}_{b}\left(z^{\prime}\right)}\right)_{{\bf \Delta}_{\textrm{mf}}[0]}^{-1}\left(\frac{\delta^{2}S_{\text{eff}}\left[{\bf \Delta},A\right]}{\delta{\bf \Delta}_{b}\left(z^{\prime}\right)\delta A_{\nu}\left(y\right)}\right)_{{\bf \Delta_{\textrm{mf}}}[0]}.
\end{align}

\section{Polarization tensor calculations\label{sec:Qmunu}}

In this appendix we provide a short discussion regarding polarization
bubbles. If $\left\{ \Phi\right\} $ collectively describes a set
of fields upon which a fermionic system depends (external electromagnetic
fields, Hubbard-Stratonovich auxiliary fields, etc), response tensors
are computed as an expansion around a reference set of values $\left\{ \overline{\Phi}\right\} $. 

\begin{align}
Q_{\Phi\Phi'}\left(x-x'\right)= & \left.\frac{\delta^{2}S_{\text{eff}}\left[\left\{ \Phi\right\} \right]}{\delta\Phi\left(x\right)\delta\Phi'\left(x^{\prime}\right)}\right|_{\left\{ \Phi\right\} =\left\{ \overline{\Phi}\right\} }\nonumber \\
= & Q_{\mathrm{bos},\Phi\Phi^{\prime}}\left(x-x^{\prime}\right)-\frac{1}{2}\left.\frac{\delta^{2}\textrm{Tr}\ln\left[-\mathcal{G}^{-1}\left[\left\{ \Phi\right\} \right]\right]}{\delta\Phi\left(x\right)\delta\Phi'\left(x^{\prime}\right)}\right|_{\left\{ \Phi\right\} =\left\{ \overline{\Phi}\right\} }\nonumber \\
= & Q_{\mathrm{bos},\Phi\Phi^{\prime}}\left(x-x^{\prime}\right)-\frac{1}{2}\left.\int d^{D}y\text{tr}\left\langle y\left|\frac{\delta\mathcal{G}\left[\left\{ \Phi\right\} \right]}{\delta\Phi\left(x\right)}\frac{\delta\mathcal{G}^{-1}\left[\left\{ \Phi\right\} \right]}{\delta\Phi'\left(x^{\prime}\right)}+\mathcal{G}\left[\left\{ \Phi\right\} \right]\frac{\delta^{2}\mathcal{G}^{-1}\left[\left\{ \Phi\right\} \right]}{\delta\Phi(x)\delta\Phi'\left(x^{\prime}\right)}\right|y\right\rangle \right|_{\left\{ \Phi\right\} =\left\{ \overline{\Phi}\right\} }\nonumber \\
= & Q_{\mathrm{bos},\Phi\Phi^{\prime}}\left(x-x^{\prime}\right)+\frac{1}{2}\left.\int d^{D}y\text{tr}\left\langle y\left|\mathcal{G}\left[\left\{ \Phi\right\} \right]\frac{\delta\mathcal{G}^{-1}\left[\left\{ \Phi\right\} \right]}{\delta\Phi\left(x\right)}\mathcal{G}\left[\left\{ \Phi\right\} \right]\frac{\delta\mathcal{G}^{-1}\left[\left\{ \Phi\right\} \right]}{\delta\Phi'\left(x^{\prime}\right)}\right|y\right\rangle \right|_{\left\{ \Phi\right\} =\left\{ \overline{\Phi}\right\} }\nonumber \\
 & -\frac{1}{2}\left.\int d^{D}y\text{tr}\left\langle y\left|\mathcal{G}\left[\left\{ \Phi\right\} \right]\frac{\delta^{2}\mathcal{G}^{-1}\left[\left\{ \Phi\right\} \right]}{\delta\Phi\left(x\right)\delta\Phi'\left(x^{\prime}\right)}\right|y\right\rangle \right|_{\left\{ \Phi\right\} =\left\{ \overline{\Phi}\right\} }.
\end{align}
Here, $Q_{\mathrm{bos},\Phi\Phi^{\prime}}\left(x-x^{\prime}\right)$
is the bosonic part of the response which arises from differentiating
the bosonic contribution to the effective action. Define real-space
vertices by

\begin{equation}
\hat{V}_{\Phi}\left(x,y,x'\right)\equiv\frac{\delta\mathcal{G}^{-1}\left[\Phi\right]\left(x,x'\right)}{\delta\Phi\left(y\right)}.
\end{equation}
The standard Green's function representation of the polarization bubbles
then follows
\begin{equation}
Q_{\Phi\Phi'}\left(x-x'\right)=Q_{\mathrm{bos},\Phi\Phi^{\prime}}\left(x-x^{\prime}\right)+\frac{1}{2}\int_{y,y',z,z'}\text{tr}\left[\mathcal{G}\left(y,z\right)\hat{V}_{\Phi}\left(z,x,z'\right)\mathcal{G}\left(z',y'\right)\hat{V}_{\Phi'}\left(y',x',y\right)\right]_{\Phi=\overline{\Phi}},\label{eq:Resp_Tens}
\end{equation}
where $\Phi$ is an arbitrary field in the system. The bare EM vertices
are defined by 
\begin{equation}
\gamma^{\mu}\left(x,y,x'\right)=\frac{\delta\mathcal{G}_{0}^{-1}\left[A\right]\left(x,x'\right)}{\delta A_{\mu}\left(y\right)}.
\end{equation}
For the models of superconductivity with a quadratic free-particle
dispersion studied in the main text, the components of the vertices
are explicitly given by

\begin{align}
\gamma^{0}\left(x,y,x'\right)= & e\tau_{3}\delta\left(x-y\right)\delta\left(x-x'\right)\\
\boldsymbol{\gamma}\left(x,y,x'\right)= & \frac{ei}{2m}\tau_{0}\left[\nabla\left(\delta\left(x-y\right)\delta\left(x-x'\right)\right)+\delta\left(x-y\right)\nabla\delta\left(x-x'\right)\right]\nonumber \\
 & +\frac{e^{2}}{m}\tau_{3}\mathbf{A}\left(x\right)\delta\left(x-y\right)\delta\left(x-x'\right),
\end{align}
and
\begin{equation}
\frac{\delta\gamma^{\nu}\left(x,y,x'\right)}{\delta A_{\mu}\left(y^{\prime}\right)}=-\frac{e^{2}}{m}\tau_{3}\delta\left(x-y^{\prime}\right)\delta\left(x-y\right)\delta\left(x-x'\right)\delta^{\mu i}\delta^{\nu j}\delta_{ij}.
\end{equation}
For the electromagnetic response, the reference value for the external
field is $A=0$. The Fourier expansion of the response is

\begin{equation}
Q^{\mu\nu}\left(x-y\right)=\int_{q}e^{-iq\cdot\left(x-y\right)}Q^{\mu\nu}\left(q\right),
\end{equation}
where $\int_{q}=TL^{d}\sum_{i\Omega_{m}}\int\frac{d{\bf q}}{\left(2\pi\right)^{d}}$.
Using the general expression in Eq.~(\ref{eq:Resp_Tens}), the undressed
polarization response is
\begin{equation}
Q^{\mu\nu}\left(q\right)=\left.\frac{1}{2}\int_{p}\text{tr}\left[\mathcal{G}\left(p+q\right)\gamma^{\mu}\left(p+q,p\right)\mathcal{G}\left(p\right)\gamma^{\nu}\left(p,p+q\right)\right]\right|_{A=0}+\frac{ne^{2}}{m}\delta^{\mu i}\delta^{\nu j}\delta_{ij}.
\end{equation}
By definition, the momentum-space vertex is defined by~\citep{Schrieffer}:
\begin{eqnarray}
\gamma^{\mu}\left(x,y,x^{\prime}\right) & = & \int_{p,q}e^{iq\left(x-y\right)}e^{ip\left(x-x^{\prime}\right)}\gamma^{\mu}\left(p+q,k\right).
\end{eqnarray}
Therefore, in the limit of zero external field, the momentum-space
vertices are~\citep{Schrieffer,Guo_Chien_He_2013}:
\begin{align}
\left.\gamma^{0}\left(p+q,p\right)\right|_{A=0} & =e\tau_{3}.\\
\left.\boldsymbol{\gamma}\left(p+q,p\right)\right|_{A=0} & =\frac{e}{m}\tau_{0}\left(\mathbf{p}+\frac{\mathbf{q}}{2}\right)=\left.\boldsymbol{\gamma}\left(p,p+q\right)\right|_{A=0}.
\end{align}
For a three-dimensional system with $p$-wave pairing, the Nambu Green's
function is

\begin{align}
\mathcal{G}\left(p\right)= & \left[i\omega_{n}-\tau_{3}\xi_{\mathbf{p}}+\Delta_{0}\left(p_{x}\tau_{1}-p_{y}\tau_{2}\right)\right]^{-1}=\frac{i\omega_{n}+\tau_{3}\xi_{\mathbf{p}}-\Delta_{0}\left(p_{x}\tau_{1}-p_{y}\tau_{2}\right)}{\left(i\omega_{n}\right)^{2}-E_{\mathbf{p}}^{2}},
\end{align}
where $\xi_{\mathbf{p}}={\bf p}^{2}/\left(2m\right)-\mu$ and $E_{\mathbf{p}}^{2}=\xi_{\mathbf{p}}^{2}+\Delta_{0}^{2}\mathbf{p}\cdot\mathbf{p}^{2}/p_{F}^{2}$.
All other bubbles appearing in the main text can be computed in a
similar fashion.

\section{Superconducting pairing in radial coordinates~\label{sec:Pairing_app}}

Here we transform the mean-field ansatz for the case of spinless $p$-wave
pairing to center-of-mass and relative coordinate representation as
a concrete example of Eqs.~(\ref{eq:SC_bos_act}) and (\ref{eq:SC_exp_act}).
The coordinate transformation is

\begin{align}
\mathbf{R} & =\frac{\mathbf{x}+\mathbf{x}^{\prime}}{2},\ \mathbf{r}=\mathbf{x}-\mathbf{x}^{\prime}.
\end{align}
The Jacobian for this transformation is unity. For a spinless fermionic
system, the $p$-wave ansatz reads
\begin{equation}
\Delta\left(\mathbf{x},\mathbf{x}^{\prime},\tau\right)=\left|\Delta\left(\frac{\mathbf{x}+\mathbf{x}^{\prime}}{2},\tau\right)\right|e^{i\Phi\left(\frac{\mathbf{x}+\mathbf{x}^{\prime}}{2},\tau\right)}\left(\partial_{x}+i\partial_{y}\right)\delta\left(\mathbf{x}-\mathbf{x}^{\prime}\right).
\end{equation}
Thus,
\begin{align}
 & \int d^{3}xd^{3}x^{\prime}\psi^{\dagger} \left(\mathbf{x},\tau\right) \Delta\left(\mathbf{x},\mathbf{x}^{\prime},\tau\right) \psi^{\dagger}\left(\mathbf{x}^{\prime},\tau\right)\nonumber \\
= & \int d^{3}R\left|\Delta\left(\mathbf{R},\tau\right)\right|e^{i\Phi\left(\mathbf{R},\tau\right)}\int d^{3}r\psi^{\dagger}\left(\mathbf{R}+\mathbf{r}/2,\tau\right)\psi^{\dagger}\left(\mathbf{R}-\mathbf{r}/2,\tau\right)\left(\partial_{r_{x}}+i\partial_{r_{y}}\right)\delta\left(\mathbf{r}\right)\nonumber \\
= & \int d^{3}R\left|\Delta\left(\mathbf{R},\tau\right)\right|e^{i\Phi\left(\mathbf{R},\tau\right)}\left[\psi^{\dagger}\left(\mathbf{R},\tau\right)\left(\partial_{R_{x}}+i\partial_{R_{y}}\right)\psi^{\dagger}\left(\mathbf{R},\tau\right)\right]\nonumber \\
= & \int d^{3}x\left|\Delta\left(\mathbf{x},\tau\right)\right|e^{i\Phi\left(\mathbf{x},\tau\right)}\left[\psi^{\dagger}\left(\mathbf{x},\tau\right)\left(\partial_{x}+i\partial_{y}\right)\psi^{\dagger}\left(\mathbf{x},\tau\right)\right],
\end{align}
after integrations by parts, identifications of gradients of fermion
fields with respect to $\mathbf{R}$ and $\mathbf{r}$ variables and
relabelling of dummy variables.

In general, non $s$-wave pairing demands a spatially dependent interaction
coefficient, say $g\left(\mathbf{x}-\mathbf{x}^{\prime}\right)$.
In this case, the $p$-wave ansatz simplifies the Gaussian part of
the identity introduced in the Hubbard-Stratonovich decomposition:
\begin{align}
 & \int d^{3}xd^{2}x^{\prime}\frac{\left|\Delta\left(\mathbf{x},\mathbf{x}^{\prime},\tau\right)\right|^{2}}{g\left(\mathbf{x}-\mathbf{x}^{\prime}\right)}\nonumber \\
= & \int d^{3}Rd^{2}r\left|\Delta\left(\mathbf{R},\tau\right)\right|e^{-i\Phi\left(\mathbf{R},\tau\right)}\left[\left(\partial_{r_{x}}-i\partial_{r_{y}}\right)\delta\left(\mathbf{r}\right)\right]g^{-1}\left(\mathbf{r}\right)\left|\Delta\left(\mathbf{R},\tau\right)\right|e^{i\Phi\left(\mathbf{R},\tau\right)}\left(\partial_{r_{x}}+i\partial_{r_{y}}\right)\delta\left(\mathbf{r}\right)\nonumber \\
= & \int d^{3}R\frac{\left|\Delta\left(\mathbf{R},\tau\right)\right|^{2}}{\widetilde{g}},
\end{align}
where we define the renormalized value for the (inverse) mass scale
of the amplitude field as
\begin{equation}
\widetilde{g}^{-1}=\int d^{2}r\left[\left(\partial_{r_{x}}-i\partial_{r_{y}}\right)\delta\left(\mathbf{r}\right)\right]g^{-1}\left(\mathbf{r}\right)\left(\partial_{r_{x}}+i\partial_{r_{y}}\right)\delta\left(\mathbf{r}\right).
\end{equation}

\section{$3\times3$ response matrix determinant calculation~\label{sec:Det_SVD}}

Here we sketch the calculation and simplification of $\det S\left(q\right)$
for the response of a charged superconductor in the presence of Coulomb,
amplitude and phase fluctuations. We first consider biasing towards
including the amplitude fluctuation effects. An expansion and consideration
of the definition in Eq.~(\ref{eq:Qbar}) returns
\begin{align}
 & \det S\left(q\right)\nonumber \\
 & =\textrm{det}\left(\begin{array}{ccc}
S^{\rho\rho} & iR^{\rho0} & iR^{\rho\beta}q_{\beta}\\
iR^{0\rho} & V^{-1}-Q^{00} & -Q^{0\beta}q_{\beta}\\
-iq_{\alpha}R^{\alpha\rho} & q_{\alpha}Q^{\alpha0} & q_{\alpha}Q^{\alpha\beta}q_{\beta}
\end{array}\right)\nonumber \\
 & =q_{\alpha}q_{\beta}S^{\rho\rho}\left[\left(V^{-1}-Q^{00}\right)\overline{Q}^{\alpha\beta}+Q^{\alpha0}Q^{0\beta}+Q^{\alpha\beta}\frac{R^{\rho0}R^{0\rho}}{S^{\rho\rho}}-Q^{0\beta}\frac{R^{\alpha\rho}R^{\rho0}}{S^{\rho\rho}}-Q^{\alpha0}\frac{R^{0\rho}R^{\rho\beta}}{S^{\rho\rho}}\right].
\end{align}
With the singular-value decomposition structure in mind, we can rework
the term in square brackets to produce
\begin{align}
\det S\left(q\right) & =q_{\alpha}q_{\beta}S^{\rho\rho}\left[\left(V^{-1}-Q^{00}\right)\overline{Q}^{\alpha\beta}+\left(\begin{array}{cc}
R^{\rho0} & Q^{\alpha0}\end{array}\right)\left(\begin{array}{cc}
\frac{Q^{\alpha\beta}}{S^{\rho\rho}} & -\frac{R^{\alpha\rho}}{S^{\rho\rho}}\\
-\frac{R^{\rho\beta}}{S^{\rho\rho}} & 1
\end{array}\right)\left(\begin{array}{c}
R^{0\rho}\\
Q^{0\beta}
\end{array}\right)\right]\nonumber \\
 & =q_{\alpha}q_{\beta}S^{\rho\rho}\left[\left(V^{-1}-\overline{Q}^{00}\right)\overline{Q}^{\alpha\beta}+\left(\begin{array}{cc}
R^{\rho0} & Q^{\alpha0}\end{array}\right)\left(\begin{array}{cc}
\frac{R^{\alpha\rho}R^{\rho\beta}}{\left(S^{\rho\rho}\right)^{2}} & -\frac{R^{\alpha\rho}}{S^{\rho\rho}}\\
-\frac{R^{\rho\beta}}{S^{\rho\rho}} & 1
\end{array}\right)\left(\begin{array}{c}
R^{0\rho}\\
Q^{0\beta}
\end{array}\right)\right].
\end{align}
Following with the decomposition, we fold the effects of Coulomb fluctuations
into $\overline{Q}^{\alpha\beta}$ to obtain

\begin{equation}
\det S\left(q\right)=S^{\rho\rho}\left(V^{-1}-\overline{Q}^{00}\right)q_{\alpha}\widetilde{\overline{Q}}^{\alpha\beta}q_{\beta}.
\end{equation}
A reversed order of the fluctuation considerations allows writing

\begin{align}
\det S\left(q\right) & =q_{\alpha}q_{\beta}\left(V^{-1}-Q^{00}\right)\left[S^{\rho\rho}\widetilde{Q}^{\alpha\beta}-R^{\alpha\rho}R^{\rho\beta}-\frac{R^{\alpha\rho}R^{\rho0}Q^{0\beta}}{\left(V^{-1}-Q^{00}\right)}-\frac{Q^{\alpha0}R^{0\rho}R^{\rho\beta}}{\left(V^{-1}-Q^{00}\right)}+\frac{Q^{\alpha\beta}R^{\rho0}R^{0\rho}}{\left(V^{-1}-Q^{00}\right)}\right].
\end{align}
Proceeding with a similar analysis, this leads to
\begin{equation}
\det S\left(q\right)=\widetilde{S}^{\rho\rho}\left(V^{-1}-Q^{00}\right)q_{\alpha}\overline{\widetilde{Q}}^{\alpha\beta}q_{\beta},
\end{equation}
proving Eq.~(\ref{eq:detS}) in the main text.

\end{widetext}

\bibliographystyle{apsrev4-1}

\begin{thebibliography}{41}%
\makeatletter
\providecommand \@ifxundefined [1]{%
 \@ifx{#1\undefined}
}%
\providecommand \@ifnum [1]{%
 \ifnum #1\expandafter \@firstoftwo
 \else \expandafter \@secondoftwo
 \fi
}%
\providecommand \@ifx [1]{%
 \ifx #1\expandafter \@firstoftwo
 \else \expandafter \@secondoftwo
 \fi
}%
\providecommand \natexlab [1]{#1}%
\providecommand \enquote  [1]{``#1''}%
\providecommand \bibnamefont  [1]{#1}%
\providecommand \bibfnamefont [1]{#1}%
\providecommand \citenamefont [1]{#1}%
\providecommand \href@noop [0]{\@secondoftwo}%
\providecommand \href [0]{\begingroup \@sanitize@url \@href}%
\providecommand \@href[1]{\@@startlink{#1}\@@href}%
\providecommand \@@href[1]{\endgroup#1\@@endlink}%
\providecommand \@sanitize@url [0]{\catcode `\\12\catcode `\$12\catcode
  `\&12\catcode `\#12\catcode `\^12\catcode `\_12\catcode `\%12\relax}%
\providecommand \@@startlink[1]{}%
\providecommand \@@endlink[0]{}%
\providecommand \url  [0]{\begingroup\@sanitize@url \@url }%
\providecommand \@url [1]{\endgroup\@href {#1}{\urlprefix }}%
\providecommand \urlprefix  [0]{URL }%
\providecommand \Eprint [0]{\href }%
\providecommand \doibase [0]{http://dx.doi.org/}%
\providecommand \selectlanguage [0]{\@gobble}%
\providecommand \bibinfo  [0]{\@secondoftwo}%
\providecommand \bibfield  [0]{\@secondoftwo}%
\providecommand \translation [1]{[#1]}%
\providecommand \BibitemOpen [0]{}%
\providecommand \bibitemStop [0]{}%
\providecommand \bibitemNoStop [0]{.\EOS\space}%
\providecommand \EOS [0]{\spacefactor3000\relax}%
\providecommand \BibitemShut  [1]{\csname bibitem#1\endcsname}%
\let\auto@bib@innerbib\@empty
\bibitem [{\citenamefont {Schrieffer}(1964)}]{Schrieffer}%
  \BibitemOpen
  \bibfield  {author} {\bibinfo {author} {\bibfnamefont {J.~R.}\ \bibnamefont
  {Schrieffer}},\ }\href@noop {} {\emph {\bibinfo {title} {Theory of
  superconductivity}}},\ \bibinfo {edition} {1st}\ ed.\ (\bibinfo  {publisher}
  {W.A. Benjamin, Inc.},\ \bibinfo {year} {1964})\BibitemShut {NoStop}%
\bibitem [{\citenamefont {Rickayzen}(1965)}]{Rickayzen}%
  \BibitemOpen
  \bibfield  {author} {\bibinfo {author} {\bibfnamefont {G.}~\bibnamefont
  {Rickayzen}},\ }\href@noop {} {\emph {\bibinfo {title} {Theory of
  superconductivity}}},\ \bibinfo {series} {Interscience monographs and texts
  in physics and astronomy}, Vol.~\bibinfo {volume} {14}\ (\bibinfo
  {publisher} {Interscience Publishers},\ \bibinfo {year} {1965})\BibitemShut
  {NoStop}%
\bibitem [{\citenamefont {Parks}(1969)}]{Parks}%
  \BibitemOpen
  \bibfield  {author} {\bibinfo {author} {\bibfnamefont {R.}~\bibnamefont
  {Parks}},\ }\href@noop {} {\emph {\bibinfo {title} {Superconductivity: Part 1
  (In Two Parts)}}},\ Superconductivity\ (\bibinfo  {publisher} {Taylor \&
  Francis},\ \bibinfo {year} {1969})\BibitemShut {NoStop}%
\bibitem [{\citenamefont {Arseev}\ \emph {et~al.}(2006)\citenamefont {Arseev},
  \citenamefont {Loiko},\ and\ \citenamefont {Fedorov}}]{Arseev_2006}%
  \BibitemOpen
  \bibfield  {author} {\bibinfo {author} {\bibfnamefont {P.~I.}\ \bibnamefont
  {Arseev}}, \bibinfo {author} {\bibfnamefont {S.~O.}\ \bibnamefont {Loiko}}, \
  and\ \bibinfo {author} {\bibfnamefont {N.~K.}\ \bibnamefont {Fedorov}},\
  }\href {http://stacks.iop.org/1063-7869/49/i=1/a=R01} {\bibfield  {journal}
  {\bibinfo  {journal} {Physics-Uspekhi}\ }\textbf {\bibinfo {volume} {49}},\
  \bibinfo {pages} {1} (\bibinfo {year} {2006})}\BibitemShut {NoStop}%
\bibitem [{\citenamefont {Anderson}\ \emph {et~al.}(2016)\citenamefont
  {Anderson}, \citenamefont {Boyack}, \citenamefont {Wu},\ and\ \citenamefont
  {Levin}}]{Anderson_2016}%
  \BibitemOpen
  \bibfield  {author} {\bibinfo {author} {\bibfnamefont {B.~M.}\ \bibnamefont
  {Anderson}}, \bibinfo {author} {\bibfnamefont {R.}~\bibnamefont {Boyack}},
  \bibinfo {author} {\bibfnamefont {C.-T.}\ \bibnamefont {Wu}}, \ and\ \bibinfo
  {author} {\bibfnamefont {K.}~\bibnamefont {Levin}},\ }\href {\doibase
  10.1103/PhysRevB.93.180504} {\bibfield  {journal} {\bibinfo  {journal} {Phys.
  Rev. B}\ }\textbf {\bibinfo {volume} {93}},\ \bibinfo {pages} {180504}
  (\bibinfo {year} {2016})}\BibitemShut {NoStop}%
\bibitem [{\citenamefont {Anderson}(1958{\natexlab{a}})}]{Anderson_1958}%
  \BibitemOpen
  \bibfield  {author} {\bibinfo {author} {\bibfnamefont {P.~W.}\ \bibnamefont
  {Anderson}},\ }\href {\doibase 10.1103/PhysRev.110.827} {\bibfield  {journal}
  {\bibinfo  {journal} {Phys. Rev.}\ }\textbf {\bibinfo {volume} {110}},\
  \bibinfo {pages} {827} (\bibinfo {year} {1958}{\natexlab{a}})}\BibitemShut
  {NoStop}%
\bibitem [{\citenamefont {Rickayzen}(1959)}]{Rickayzen_1959}%
  \BibitemOpen
  \bibfield  {author} {\bibinfo {author} {\bibfnamefont {G.}~\bibnamefont
  {Rickayzen}},\ }\href {\doibase 10.1103/PhysRev.115.795} {\bibfield
  {journal} {\bibinfo  {journal} {Phys. Rev.}\ }\textbf {\bibinfo {volume}
  {115}},\ \bibinfo {pages} {795} (\bibinfo {year} {1959})}\BibitemShut
  {NoStop}%
\bibitem [{\citenamefont {Nambu}(1960)}]{Nambu_1960}%
  \BibitemOpen
  \bibfield  {author} {\bibinfo {author} {\bibfnamefont {Y.}~\bibnamefont
  {Nambu}},\ }\href {\doibase 10.1103/PhysRev.117.648} {\bibfield  {journal}
  {\bibinfo  {journal} {Phys. Rev.}\ }\textbf {\bibinfo {volume} {117}},\
  \bibinfo {pages} {648} (\bibinfo {year} {1960})}\BibitemShut {NoStop}%
\bibitem [{\citenamefont {Pekker}\ and\ \citenamefont
  {Varma}(2015)}]{Pekker_Varma_2015}%
  \BibitemOpen
  \bibfield  {author} {\bibinfo {author} {\bibfnamefont {D.}~\bibnamefont
  {Pekker}}\ and\ \bibinfo {author} {\bibfnamefont {C.}~\bibnamefont {Varma}},\
  }\href {\doibase 10.1146/annurev-conmatphys-031214-014350} {\bibfield
  {journal} {\bibinfo  {journal} {Annual Review of Condensed Matter Physics}\
  }\textbf {\bibinfo {volume} {6}},\ \bibinfo {pages} {269} (\bibinfo {year}
  {2015})}\BibitemShut {NoStop}%
\bibitem [{\citenamefont {Pollet}\ and\ \citenamefont
  {Prokof'ev}(2012)}]{Pollet_Prokofev_2012}%
  \BibitemOpen
  \bibfield  {author} {\bibinfo {author} {\bibfnamefont {L.}~\bibnamefont
  {Pollet}}\ and\ \bibinfo {author} {\bibfnamefont {N.}~\bibnamefont
  {Prokof'ev}},\ }\href {\doibase 10.1103/PhysRevLett.109.010401} {\bibfield
  {journal} {\bibinfo  {journal} {Phys. Rev. Lett.}\ }\textbf {\bibinfo
  {volume} {109}},\ \bibinfo {pages} {010401} (\bibinfo {year}
  {2012})}\BibitemShut {NoStop}%
\bibitem [{\citenamefont {Endres}\ \emph {et~al.}(2012)\citenamefont {Endres},
  \citenamefont {Fukuhara}, \citenamefont {Pekker}, \citenamefont {Cheneau},
  \citenamefont {Schau$\beta$}, \citenamefont {Gross}, \citenamefont {Demler},
  \citenamefont {Kuhr},\ and\ \citenamefont {Bloch}}]{Bloch_2012}%
  \BibitemOpen
  \bibfield  {author} {\bibinfo {author} {\bibfnamefont {M.}~\bibnamefont
  {Endres}}, \bibinfo {author} {\bibfnamefont {T.}~\bibnamefont {Fukuhara}},
  \bibinfo {author} {\bibfnamefont {D.}~\bibnamefont {Pekker}}, \bibinfo
  {author} {\bibfnamefont {M.}~\bibnamefont {Cheneau}}, \bibinfo {author}
  {\bibfnamefont {P.}~\bibnamefont {Schau$\beta$}}, \bibinfo {author}
  {\bibfnamefont {C.}~\bibnamefont {Gross}}, \bibinfo {author} {\bibfnamefont
  {E.}~\bibnamefont {Demler}}, \bibinfo {author} {\bibfnamefont
  {S.}~\bibnamefont {Kuhr}}, \ and\ \bibinfo {author} {\bibfnamefont
  {I.}~\bibnamefont {Bloch}},\ }\href {\doibase 10.1038/nature11255} {\bibfield
   {journal} {\bibinfo  {journal} {Nature}\ }\textbf {\bibinfo {volume}
  {487}},\ \bibinfo {pages} {454} (\bibinfo {year} {2012})}\BibitemShut
  {NoStop}%
\bibitem [{\citenamefont {Sherman}\ \emph {et~al.}(2015)\citenamefont
  {Sherman}, \citenamefont {Pracht}, \citenamefont {Gorshunov}, \citenamefont
  {Poran}, \citenamefont {Jesudasan}, \citenamefont {Chand}, \citenamefont
  {Raychaudhuri}, \citenamefont {Swanson}, \citenamefont {Trivedi},
  \citenamefont {Auerbach}, \citenamefont {Scheffler}, \citenamefont
  {Frydman},\ and\ \citenamefont {Dressel}}]{Sherman_2015}%
  \BibitemOpen
  \bibfield  {author} {\bibinfo {author} {\bibfnamefont {D.}~\bibnamefont
  {Sherman}}, \bibinfo {author} {\bibfnamefont {U.~S.}\ \bibnamefont {Pracht}},
  \bibinfo {author} {\bibfnamefont {B.}~\bibnamefont {Gorshunov}}, \bibinfo
  {author} {\bibfnamefont {S.}~\bibnamefont {Poran}}, \bibinfo {author}
  {\bibfnamefont {J.}~\bibnamefont {Jesudasan}}, \bibinfo {author}
  {\bibfnamefont {M.}~\bibnamefont {Chand}}, \bibinfo {author} {\bibfnamefont
  {P.}~\bibnamefont {Raychaudhuri}}, \bibinfo {author} {\bibfnamefont
  {M.}~\bibnamefont {Swanson}}, \bibinfo {author} {\bibfnamefont
  {N.}~\bibnamefont {Trivedi}}, \bibinfo {author} {\bibfnamefont
  {A.}~\bibnamefont {Auerbach}}, \bibinfo {author} {\bibfnamefont
  {M.}~\bibnamefont {Scheffler}}, \bibinfo {author} {\bibfnamefont
  {A.}~\bibnamefont {Frydman}}, \ and\ \bibinfo {author} {\bibfnamefont
  {M.}~\bibnamefont {Dressel}},\ }\href {\doibase 10.1038/nphys3227} {\bibfield
   {journal} {\bibinfo  {journal} {Nature Physics}\ }\textbf {\bibinfo {volume}
  {11}},\ \bibinfo {pages} {188} (\bibinfo {year} {2015})}\BibitemShut
  {NoStop}%
\bibitem [{\citenamefont {Littlewood}\ and\ \citenamefont
  {Varma}(1981)}]{Littlewood_Varma_1981}%
  \BibitemOpen
  \bibfield  {author} {\bibinfo {author} {\bibfnamefont {P.~B.}\ \bibnamefont
  {Littlewood}}\ and\ \bibinfo {author} {\bibfnamefont {C.~M.}\ \bibnamefont
  {Varma}},\ }\href {\doibase 10.1103/PhysRevLett.47.811} {\bibfield  {journal}
  {\bibinfo  {journal} {Phys. Rev. Lett.}\ }\textbf {\bibinfo {volume} {47}},\
  \bibinfo {pages} {811} (\bibinfo {year} {1981})}\BibitemShut {NoStop}%
\bibitem [{\citenamefont {Browne}\ and\ \citenamefont
  {Levin}(1983)}]{Browne_Levin_1983}%
  \BibitemOpen
  \bibfield  {author} {\bibinfo {author} {\bibfnamefont {D.~A.}\ \bibnamefont
  {Browne}}\ and\ \bibinfo {author} {\bibfnamefont {K.}~\bibnamefont {Levin}},\
  }\href {\doibase 10.1103/PhysRevB.28.4029} {\bibfield  {journal} {\bibinfo
  {journal} {Phys. Rev. B}\ }\textbf {\bibinfo {volume} {28}},\ \bibinfo
  {pages} {4029} (\bibinfo {year} {1983})}\BibitemShut {NoStop}%
\bibitem [{\citenamefont {Matsunaga}\ \emph {et~al.}(2014)\citenamefont
  {Matsunaga}, \citenamefont {Tsuji}, \citenamefont {Fujita}, \citenamefont
  {Sugioka}, \citenamefont {Makise}, \citenamefont {Uzawa}, \citenamefont
  {Terai}, \citenamefont {Wang}, \citenamefont {Aoki},\ and\ \citenamefont
  {Shimano}}]{Matsunaga1145}%
  \BibitemOpen
  \bibfield  {author} {\bibinfo {author} {\bibfnamefont {R.}~\bibnamefont
  {Matsunaga}}, \bibinfo {author} {\bibfnamefont {N.}~\bibnamefont {Tsuji}},
  \bibinfo {author} {\bibfnamefont {H.}~\bibnamefont {Fujita}}, \bibinfo
  {author} {\bibfnamefont {A.}~\bibnamefont {Sugioka}}, \bibinfo {author}
  {\bibfnamefont {K.}~\bibnamefont {Makise}}, \bibinfo {author} {\bibfnamefont
  {Y.}~\bibnamefont {Uzawa}}, \bibinfo {author} {\bibfnamefont
  {H.}~\bibnamefont {Terai}}, \bibinfo {author} {\bibfnamefont
  {Z.}~\bibnamefont {Wang}}, \bibinfo {author} {\bibfnamefont {H.}~\bibnamefont
  {Aoki}}, \ and\ \bibinfo {author} {\bibfnamefont {R.}~\bibnamefont
  {Shimano}},\ }\href {\doibase 10.1126/science.1254697} {\bibfield  {journal}
  {\bibinfo  {journal} {Science}\ }\textbf {\bibinfo {volume} {345}},\ \bibinfo
  {pages} {1145} (\bibinfo {year} {2014})}\BibitemShut {NoStop}%
\bibitem [{\citenamefont {Shimano}\ and\ \citenamefont
  {Tsuji}(2019)}]{shimano_higgs}%
  \BibitemOpen
  \bibfield  {author} {\bibinfo {author} {\bibfnamefont {R.}~\bibnamefont
  {Shimano}}\ and\ \bibinfo {author} {\bibfnamefont {N.}~\bibnamefont
  {Tsuji}},\ }\href@noop {} {\bibfield  {journal} {\bibinfo  {journal}
  {arXiv:1906.09401}\ } (\bibinfo {year} {2019})}\BibitemShut {NoStop}%
\bibitem [{\citenamefont {Fradkin}\ \emph {et~al.}(2015)\citenamefont
  {Fradkin}, \citenamefont {Kivelson},\ and\ \citenamefont
  {Tranquada}}]{RevModPhys.87.457}%
  \BibitemOpen
  \bibfield  {author} {\bibinfo {author} {\bibfnamefont {E.}~\bibnamefont
  {Fradkin}}, \bibinfo {author} {\bibfnamefont {S.~A.}\ \bibnamefont
  {Kivelson}}, \ and\ \bibinfo {author} {\bibfnamefont {J.~M.}\ \bibnamefont
  {Tranquada}},\ }\href {\doibase 10.1103/RevModPhys.87.457} {\bibfield
  {journal} {\bibinfo  {journal} {Rev. Mod. Phys.}\ }\textbf {\bibinfo {volume}
  {87}},\ \bibinfo {pages} {457} (\bibinfo {year} {2015})}\BibitemShut
  {NoStop}%
\bibitem [{\citenamefont {Zha}\ \emph {et~al.}(1995)\citenamefont {Zha},
  \citenamefont {Levin},\ and\ \citenamefont {Liu}}]{Zha_1995}%
  \BibitemOpen
  \bibfield  {author} {\bibinfo {author} {\bibfnamefont {Y.}~\bibnamefont
  {Zha}}, \bibinfo {author} {\bibfnamefont {K.}~\bibnamefont {Levin}}, \ and\
  \bibinfo {author} {\bibfnamefont {D.~Z.}\ \bibnamefont {Liu}},\ }\href
  {\doibase 10.1103/PhysRevB.51.6602} {\bibfield  {journal} {\bibinfo
  {journal} {Phys. Rev. B}\ }\textbf {\bibinfo {volume} {51}},\ \bibinfo
  {pages} {6602} (\bibinfo {year} {1995})}\BibitemShut {NoStop}%
\bibitem [{\citenamefont {Nozi\'eres}\ and\ \citenamefont
  {Pines}(1990)}]{PinesNozieres2}%
  \BibitemOpen
  \bibfield  {author} {\bibinfo {author} {\bibfnamefont {P.}~\bibnamefont
  {Nozi\'eres}}\ and\ \bibinfo {author} {\bibfnamefont {D.}~\bibnamefont
  {Pines}},\ }\href@noop {} {\emph {\bibinfo {title} {Theory of Quantum Liquids
  Vol. II}}}\ (\bibinfo  {publisher} {Addison-Wesley, California},\ \bibinfo
  {year} {1990})\BibitemShut {NoStop}%
\bibitem [{\citenamefont {Martin}(1967)}]{Martin_1967}%
  \BibitemOpen
  \bibfield  {author} {\bibinfo {author} {\bibfnamefont {P.~C.}\ \bibnamefont
  {Martin}},\ }\href {\doibase 10.1103/PhysRev.161.143} {\bibfield  {journal}
  {\bibinfo  {journal} {Phys. Rev.}\ }\textbf {\bibinfo {volume} {161}},\
  \bibinfo {pages} {143} (\bibinfo {year} {1967})}\BibitemShut {NoStop}%
\bibitem [{\citenamefont {Millis}(1987)}]{Millis_1987}%
  \BibitemOpen
  \bibfield  {author} {\bibinfo {author} {\bibfnamefont {A.~J.}\ \bibnamefont
  {Millis}},\ }\href {\doibase 10.1103/PhysRevB.35.151} {\bibfield  {journal}
  {\bibinfo  {journal} {Phys. Rev. B}\ }\textbf {\bibinfo {volume} {35}},\
  \bibinfo {pages} {151} (\bibinfo {year} {1987})}\BibitemShut {NoStop}%
\bibitem [{\citenamefont {Boyack}\ \emph {et~al.}(2017)\citenamefont {Boyack},
  \citenamefont {Wu}, \citenamefont {Anderson},\ and\ \citenamefont
  {Levin}}]{Boyack_2017}%
  \BibitemOpen
  \bibfield  {author} {\bibinfo {author} {\bibfnamefont {R.}~\bibnamefont
  {Boyack}}, \bibinfo {author} {\bibfnamefont {C.-T.}\ \bibnamefont {Wu}},
  \bibinfo {author} {\bibfnamefont {B.~M.}\ \bibnamefont {Anderson}}, \ and\
  \bibinfo {author} {\bibfnamefont {K.}~\bibnamefont {Levin}},\ }\href
  {\doibase 10.1103/PhysRevB.95.214501} {\bibfield  {journal} {\bibinfo
  {journal} {Phys. Rev. B}\ }\textbf {\bibinfo {volume} {95}},\ \bibinfo
  {pages} {214501} (\bibinfo {year} {2017})}\BibitemShut {NoStop}%
\bibitem [{\citenamefont {Hoyos}\ \emph {et~al.}(2014)\citenamefont {Hoyos},
  \citenamefont {Moroz},\ and\ \citenamefont {Son}}]{Hoyos_Moroz_Son_2014}%
  \BibitemOpen
  \bibfield  {author} {\bibinfo {author} {\bibfnamefont {C.}~\bibnamefont
  {Hoyos}}, \bibinfo {author} {\bibfnamefont {S.}~\bibnamefont {Moroz}}, \ and\
  \bibinfo {author} {\bibfnamefont {D.~T.}\ \bibnamefont {Son}},\ }\href
  {\doibase 10.1103/PhysRevB.89.174507} {\bibfield  {journal} {\bibinfo
  {journal} {Phys. Rev. B}\ }\textbf {\bibinfo {volume} {89}},\ \bibinfo
  {pages} {174507} (\bibinfo {year} {2014})}\BibitemShut {NoStop}%
\bibitem [{\citenamefont {Ariad}\ \emph {et~al.}(2015)\citenamefont {Ariad},
  \citenamefont {Grosfeld},\ and\ \citenamefont {Seradjeh}}]{Ariad_2015}%
  \BibitemOpen
  \bibfield  {author} {\bibinfo {author} {\bibfnamefont {D.}~\bibnamefont
  {Ariad}}, \bibinfo {author} {\bibfnamefont {E.}~\bibnamefont {Grosfeld}}, \
  and\ \bibinfo {author} {\bibfnamefont {B.}~\bibnamefont {Seradjeh}},\ }\href
  {\doibase 10.1103/PhysRevB.92.035136} {\bibfield  {journal} {\bibinfo
  {journal} {Phys. Rev. B}\ }\textbf {\bibinfo {volume} {92}},\ \bibinfo
  {pages} {035136} (\bibinfo {year} {2015})}\BibitemShut {NoStop}%
\bibitem [{\citenamefont {Lutchyn}\ \emph {et~al.}(2008)\citenamefont
  {Lutchyn}, \citenamefont {Nagornykh},\ and\ \citenamefont
  {Yakovenko}}]{Yakovenko_2008}%
  \BibitemOpen
  \bibfield  {author} {\bibinfo {author} {\bibfnamefont {R.~M.}\ \bibnamefont
  {Lutchyn}}, \bibinfo {author} {\bibfnamefont {P.}~\bibnamefont {Nagornykh}},
  \ and\ \bibinfo {author} {\bibfnamefont {V.~M.}\ \bibnamefont {Yakovenko}},\
  }\href {\doibase 10.1103/PhysRevB.77.144516} {\bibfield  {journal} {\bibinfo
  {journal} {Phys. Rev. B}\ }\textbf {\bibinfo {volume} {77}},\ \bibinfo
  {pages} {144516} (\bibinfo {year} {2008})}\BibitemShut {NoStop}%
\bibitem [{\citenamefont {Lutchyn}\ \emph {et~al.}(2009)\citenamefont
  {Lutchyn}, \citenamefont {Nagornykh},\ and\ \citenamefont
  {Yakovenko}}]{Yakoveko_2009}%
  \BibitemOpen
  \bibfield  {author} {\bibinfo {author} {\bibfnamefont {R.~M.}\ \bibnamefont
  {Lutchyn}}, \bibinfo {author} {\bibfnamefont {P.}~\bibnamefont {Nagornykh}},
  \ and\ \bibinfo {author} {\bibfnamefont {V.~M.}\ \bibnamefont {Yakovenko}},\
  }\href {\doibase 10.1103/PhysRevB.80.104508} {\bibfield  {journal} {\bibinfo
  {journal} {Phys. Rev. B}\ }\textbf {\bibinfo {volume} {80}},\ \bibinfo
  {pages} {104508} (\bibinfo {year} {2009})}\BibitemShut {NoStop}%
\bibitem [{\citenamefont {Abrikosov}\ \emph {et~al.}(1965)\citenamefont
  {Abrikosov}, \citenamefont {Gorkov},\ and\ \citenamefont
  {Dzyaloshinskii}}]{AGDBook}%
  \BibitemOpen
  \bibfield  {author} {\bibinfo {author} {\bibfnamefont {A.~A.}\ \bibnamefont
  {Abrikosov}}, \bibinfo {author} {\bibfnamefont {L.~P.}\ \bibnamefont
  {Gorkov}}, \ and\ \bibinfo {author} {\bibfnamefont {I.~E.}\ \bibnamefont
  {Dzyaloshinskii}},\ }\href@noop {} {\emph {\bibinfo {title} {Quantum field
  theoretical methods in statistical physics}}},\ \bibinfo {edition} {2nd}\
  ed.,\ Series in natural philosophy\ (\bibinfo  {publisher} {Pergamon press},\
  \bibinfo {year} {1965})\BibitemShut {NoStop}%
\bibitem [{\citenamefont {Guo}\ \emph {et~al.}(2013)\citenamefont {Guo},
  \citenamefont {Chien},\ and\ \citenamefont {He}}]{Guo_Chien_He_2013}%
  \BibitemOpen
  \bibfield  {author} {\bibinfo {author} {\bibfnamefont {H.}~\bibnamefont
  {Guo}}, \bibinfo {author} {\bibfnamefont {C.-C.}\ \bibnamefont {Chien}}, \
  and\ \bibinfo {author} {\bibfnamefont {Y.}~\bibnamefont {He}},\ }\href
  {\doibase 10.1007/s10909-012-0853-7} {\bibfield  {journal} {\bibinfo
  {journal} {Journal of Low Temperature Physics}\ }\textbf {\bibinfo {volume}
  {172}},\ \bibinfo {pages} {5} (\bibinfo {year} {2013})}\BibitemShut {NoStop}%
\bibitem [{\citenamefont {Nozi\'eres}\ and\ \citenamefont
  {Pines}(1966)}]{PinesNozieres1}%
  \BibitemOpen
  \bibfield  {author} {\bibinfo {author} {\bibfnamefont {P.}~\bibnamefont
  {Nozi\'eres}}\ and\ \bibinfo {author} {\bibfnamefont {D.}~\bibnamefont
  {Pines}},\ }\href@noop {} {\emph {\bibinfo {title} {Theory of Quantum Liquids
  Vol. I}}}\ (\bibinfo  {publisher} {W.A. Benjamin},\ \bibinfo {year}
  {1966})\BibitemShut {NoStop}%
\bibitem [{\citenamefont {Sharapov}\ \emph {et~al.}(2001)\citenamefont
  {Sharapov}, \citenamefont {Beck},\ and\ \citenamefont
  {Loktev}}]{Sharapov_2001}%
  \BibitemOpen
  \bibfield  {author} {\bibinfo {author} {\bibfnamefont {S.~G.}\ \bibnamefont
  {Sharapov}}, \bibinfo {author} {\bibfnamefont {H.}~\bibnamefont {Beck}}, \
  and\ \bibinfo {author} {\bibfnamefont {V.~M.}\ \bibnamefont {Loktev}},\
  }\href {\doibase 10.1103/PhysRevB.64.134519} {\bibfield  {journal} {\bibinfo
  {journal} {Phys. Rev. B}\ }\textbf {\bibinfo {volume} {64}},\ \bibinfo
  {pages} {134519} (\bibinfo {year} {2001})}\BibitemShut {NoStop}%
\bibitem [{\citenamefont {Kulik}\ \emph {et~al.}(1981)\citenamefont {Kulik},
  \citenamefont {Entin-Wohlman},\ and\ \citenamefont {Orbach}}]{Kulik1981}%
  \BibitemOpen
  \bibfield  {author} {\bibinfo {author} {\bibfnamefont {I.~O.}\ \bibnamefont
  {Kulik}}, \bibinfo {author} {\bibfnamefont {O.}~\bibnamefont
  {Entin-Wohlman}}, \ and\ \bibinfo {author} {\bibfnamefont {R.}~\bibnamefont
  {Orbach}},\ }\href {\doibase 10.1007/BF00115617} {\bibfield  {journal}
  {\bibinfo  {journal} {Journal of Low Temperature Physics}\ }\textbf {\bibinfo
  {volume} {43}},\ \bibinfo {pages} {591} (\bibinfo {year} {1981})}\BibitemShut
  {NoStop}%
\bibitem [{Note1()}]{Note1}%
  \BibitemOpen
  \bibinfo {note} {We assume that the procedure of integrating out the fermions
  is well defined.}\BibitemShut {Stop}%
\bibitem [{\citenamefont {Ambegaokar}\ and\ \citenamefont
  {Kadanoff}(1961)}]{Ambegaokar_Kadanoff_1961}%
  \BibitemOpen
  \bibfield  {author} {\bibinfo {author} {\bibfnamefont {V.}~\bibnamefont
  {Ambegaokar}}\ and\ \bibinfo {author} {\bibfnamefont {L.~P.}\ \bibnamefont
  {Kadanoff}},\ }\href {\doibase 10.1007/BF02787879} {\bibfield  {journal}
  {\bibinfo  {journal} {II Nuovo Cimento}\ }\textbf {\bibinfo {volume} {22}},\
  \bibinfo {pages} {914} (\bibinfo {year} {1961})}\BibitemShut {NoStop}%
\bibitem [{\citenamefont {Anderson}(1958{\natexlab{b}})}]{Anderson_1958b}%
  \BibitemOpen
  \bibfield  {author} {\bibinfo {author} {\bibfnamefont {P.~W.}\ \bibnamefont
  {Anderson}},\ }\href {\doibase 10.1103/PhysRev.112.1900} {\bibfield
  {journal} {\bibinfo  {journal} {Phys. Rev.}\ }\textbf {\bibinfo {volume}
  {112}},\ \bibinfo {pages} {1900} (\bibinfo {year}
  {1958}{\natexlab{b}})}\BibitemShut {NoStop}%
\bibitem [{\citenamefont {Bogoljubov}\ \emph {et~al.}(1958)\citenamefont
  {Bogoljubov}, \citenamefont {Tolmachov},\ and\ \citenamefont
  {$\check{\text{S}}$irkov}}]{Bogoliubov1958}%
  \BibitemOpen
  \bibfield  {author} {\bibinfo {author} {\bibfnamefont {N.~N.}\ \bibnamefont
  {Bogoljubov}}, \bibinfo {author} {\bibfnamefont {V.~V.}\ \bibnamefont
  {Tolmachov}}, \ and\ \bibinfo {author} {\bibfnamefont {D.~V.}\ \bibnamefont
  {$\check{\text{S}}$irkov}},\ }\href {\doibase 10.1002/prop.19580061102}
  {\bibfield  {journal} {\bibinfo  {journal} {Fortschritte der Physik}\
  }\textbf {\bibinfo {volume} {6}},\ \bibinfo {pages} {605} (\bibinfo {year}
  {1958})}\BibitemShut {NoStop}%
\bibitem [{\citenamefont {Bogoljubov}\ \emph {et~al.}(1959)\citenamefont
  {Bogoljubov}, \citenamefont {Tolmachov},\ and\ \citenamefont
  {$\check{\text{S}}$irkov}}]{BogoliubovBook}%
  \BibitemOpen
  \bibfield  {author} {\bibinfo {author} {\bibfnamefont {N.~N.}\ \bibnamefont
  {Bogoljubov}}, \bibinfo {author} {\bibfnamefont {V.~V.}\ \bibnamefont
  {Tolmachov}}, \ and\ \bibinfo {author} {\bibfnamefont {D.~V.}\ \bibnamefont
  {$\check{\text{S}}$irkov}},\ }\href@noop {} {\emph {\bibinfo {title} {A New
  Method in the Theory of Superconductivity}}}\ (\bibinfo  {publisher}
  {Consultants Buerau},\ \bibinfo {year} {1959})\BibitemShut {NoStop}%
\bibitem [{\citenamefont {Garate}(2013)}]{Garate_2013}%
  \BibitemOpen
  \bibfield  {author} {\bibinfo {author} {\bibfnamefont {I.}~\bibnamefont
  {Garate}},\ }\href {\doibase 10.1103/PhysRevB.88.094511} {\bibfield
  {journal} {\bibinfo  {journal} {Phys. Rev. B}\ }\textbf {\bibinfo {volume}
  {88}},\ \bibinfo {pages} {094511} (\bibinfo {year} {2013})}\BibitemShut
  {NoStop}%
\bibitem [{\citenamefont {Carlson}\ and\ \citenamefont
  {Goldman}(1973)}]{Carlson_Goldman_1973}%
  \BibitemOpen
  \bibfield  {author} {\bibinfo {author} {\bibfnamefont {R.~V.}\ \bibnamefont
  {Carlson}}\ and\ \bibinfo {author} {\bibfnamefont {A.~M.}\ \bibnamefont
  {Goldman}},\ }\href {\doibase 10.1103/PhysRevLett.31.880} {\bibfield
  {journal} {\bibinfo  {journal} {Phys. Rev. Lett.}\ }\textbf {\bibinfo
  {volume} {31}},\ \bibinfo {pages} {880} (\bibinfo {year} {1973})}\BibitemShut
  {NoStop}%
\bibitem [{\citenamefont {Carlson}\ and\ \citenamefont
  {Goldman}(1975)}]{Carlson_Goldman_1975}%
  \BibitemOpen
  \bibfield  {author} {\bibinfo {author} {\bibfnamefont {R.~V.}\ \bibnamefont
  {Carlson}}\ and\ \bibinfo {author} {\bibfnamefont {A.~M.}\ \bibnamefont
  {Goldman}},\ }\href {\doibase 10.1103/PhysRevLett.34.11} {\bibfield
  {journal} {\bibinfo  {journal} {Phys. Rev. Lett.}\ }\textbf {\bibinfo
  {volume} {34}},\ \bibinfo {pages} {11} (\bibinfo {year} {1975})}\BibitemShut
  {NoStop}%
\bibitem [{\citenamefont {Kosztin}\ \emph {et~al.}(2000)\citenamefont
  {Kosztin}, \citenamefont {Chen}, \citenamefont {Kao},\ and\ \citenamefont
  {Levin}}]{Levin_2000}%
  \BibitemOpen
  \bibfield  {author} {\bibinfo {author} {\bibfnamefont {I.}~\bibnamefont
  {Kosztin}}, \bibinfo {author} {\bibfnamefont {Q.}~\bibnamefont {Chen}},
  \bibinfo {author} {\bibfnamefont {Y.-J.}\ \bibnamefont {Kao}}, \ and\
  \bibinfo {author} {\bibfnamefont {K.}~\bibnamefont {Levin}},\ }\href
  {\doibase 10.1103/PhysRevB.61.11662} {\bibfield  {journal} {\bibinfo
  {journal} {Phys. Rev. B}\ }\textbf {\bibinfo {volume} {61}},\ \bibinfo
  {pages} {11662} (\bibinfo {year} {2000})}\BibitemShut {NoStop}%
\bibitem [{Note2()}]{Note2}%
  \BibitemOpen
  \bibinfo {note} {In Ref.~\protect \citep {Millis_1987} the prefactor of $1/m$
  in the EM vertices was omitted. Furthermore, the explicit form for $dS_{p}$
  was unspecified. The definition given in Ref.~\protect \citep {Millis_1987}
  is that, up to a constant, it is the angle-dependent density of states. Using
  this definition, we accordingly find $\DOTSI \intop \ilimits@ dS_{p}\protect
  \qopname \relax o{sin}^{2}\left (\theta \right )f\left (\theta \right )\sim
  \protect \frac {mp_{F}}{\pi ^{2}}p_{F}^{2}\DOTSI \intop \ilimits@ \protect
  \frac {d\theta d\phi }{4\pi }\protect \qopname \relax o{sin}^{3}\left (\theta
  \right )f\left (\theta \right )=\protect \frac {3}{2}\protect \frac
  {n}{m}m^{2}\DOTSI \intop \ilimits@ dx\left (1-x^{2}\right )f\left (x\right
  )$. The factor of $m^{2}$ drops out once the vertices are appropriately
  restored. There is an additional factor of $\pi $ that also must be restored.
  Nevertheless, our result is in exact agreement with Eq.~(37.15) of
  Ref.~\protect \citep {AGDBook} for the $s$-wave case (accounting for the
  differences in definition of response kernel.).}\BibitemShut {Stop}%
\end{thebibliography}

\end{document}